\documentclass[preprint, superscriptaddress]{revtex4-2}

\makeatletter
\renewcommand{\p@subsection}{}
\renewcommand{\p@subsubsection}{}
\makeatother
\usepackage{graphicx}

\newcommand{\NIMS}{National Institute for Materials Science, 1-2-1 Sengen, Tsukuba, Japan}
\newcommand{\UT}{University of Tsukuba, 1-1-1 Tennoidai, Tsukuba, Japan}

\begin{document}
\title{XERUS: \\ An open-source tool for quick XRD phase identification and refinement automation}

\author{Pedro Baptista de Castro}
\email{CASTRO.Pedro@nims.go.jp}
\affiliation{\NIMS}
\affiliation{\UT}
\author{Kensei Terashima}
\email{TERASHIMA.Kensei@nims.go.jp}
\affiliation{\NIMS}
\author{Miren Garbine Esparza Echevarria}
\affiliation{\NIMS}
\affiliation{\UT}
\author{Hiroyuki Takeya}
\affiliation{\NIMS}
\author{Yoshihiko Takano}
\affiliation{\NIMS}
\affiliation{\UT}


\begin{abstract}
  Analysis of XRD diffraction patterns is one of the keystones of materials science and materials research. With the advancement of data-driven methods for materials design, candidate materials can be quickly screened for the study of a desiredphysical property. Efficient methods to automatically analyze and identify phases present in a given pattern, are paramount for the success of this new paradigm. To aid this process, the open source python package Xray Estimation and Refinement Using Similarity (XERUS)  for semi-automatic/automatic phase identification is presented. XERUS takes advantages of open crystal structure databases, not relying on proprietary databases, to obtain crystal structures on the fly, being then chemical space agnostic. By wrapping around GSASII, it can automatically simulate patterns and calculate similarity measures used for phase identification. Our approach is simple and quick but also applicable to multiphase identification, by coupling the similarity calculations with quick refinements followed by an iterative peak removal process. XERUS is shown in action in four different experimental datasets and also it is benchmarked against a recently proposed deep learning method for a mixture dataset covering the Li-Mn-O-F chemical space. XERUS will be freely available on https://www.github.com/pedrobcst/Xerus/
\end{abstract}
\maketitle

\section{Introduction}
The establishment of high through-put computational databases such as the Materials Project (MP)\cite{materialsprojectJain2013}, 
the Open Quantum Materials Database\cite{oqmdkirklin2015open} (OQMD), and AFLOW\cite{aflowcurtarolo2012aflow}, ignited a new age in materials science, where the conventional trial-and-error approach for materials design is replaced by a data-centric approach. More recently, artificial intelligence-based  methods are also becoming a key driver in this new paradigm of materials design\cite{balachandran2019machine}, allowing not only the creation of models for physical property prediction with machine learning tools, but also allowing the extraction and creation of experimental datasets\cite{olivetti2020data,Luca2021-Supermat}. \\
These new tools, that allow for the quick screening of candidate materials for synthesis and experimental evaluation, also require
methods than can quickly identify the synthesis success of the proposed candidate. This often involves analyzing the
Xray Diffraction (XRD) of the obtained materials, a process that is currently one of the primary bottlenecks of 
data-driven materials search and automation since it requires time consuming human analysis\cite{szymanski2021toward}. \\
To solve this issue, several methods for automating XRD analysis has been proposed, with most recent approaches
focusing in the use of deep-learning (DL) assisted models. For example, models for automatic crystal system and space group classification\cite{tiong2020identification,oviedo2019fast},
single phase\cite{wang2020rapid,lee2020deep} and possible multiphase\cite{lee2021data,szymanski2021probabilistic} have been proposed. Here we highlight as an example, the recent work of Syzmanski\cite{szymanski2021probabilistic} \textit{et. al}, where the entire process of simulation and model building  for multiphase classification of the Li-Mn-Ti-O-F chemical space took approximately 20 hours of computational time, 
achieving an accuracy of 92\% for multiphase classification. However, these models\cite{oviedo2019fast,wang2020rapid,lee2020deep, lee2021data, szymanski2021probabilistic} usually rely on commercial databases such as the ICSD\cite{hellenbrandt2004inorganic}.Evenmore, they require simulation of large synthetic datasets for model training which are both time consuming and restricts the model to a certain chemical space. On the other hand, candidate-based materials search often involves different materials spaning a large chemical space, where sometimes even the optimal synthesis route can be unknown. 
For example, in the search of new magnetocaloric materials\cite{Castro2020-12,Bocarsly2017,court2021inverse}, both the work of Bocarsly\cite{Bocarsly2017} \textit{et. al}
and Court\cite{court2021inverse} \textit{et. al}, all the identified candidate materials belonged to different chemical spaces, ie. Mn-Nb-S and Co-Sc-V. Moreover, in the recent work of Xiong\cite{xiong2021optimizing} \textit{et. al}, by high through-put screening of the MP database for efficient photocatalysts, 11 different materials were identified and synthesized for experimental evaluation, 
all belonging to different chemical spaces, such as Ca-Pb-O, Ba-Nb-Mn-O, Sr-In-O, Ca-In-O, etc. In these scenarios, the use of DL assisted methods for XRD classification, will drastically increase the preparation time, since large synthetic datasets followed by model training would be necessary for each individual chemical space, slowing down the whole data-driven search process. \\
Thus, there is a need for an automatic/semi-automatic tool, that allows quick identification/suggestion of possible synthesis results, which can adapt on-the-fly to the chemical space. In this work, we introduce an alternative approach for phase identification, by combining similarity calculation of X-ray patterns with refinements, which does not require any pre-trained model, in the open-source python framework Xray Estimation and Refinement Using Similarity (XERUS) as an attempt to solve this problem. \\
XERUS takes advantage of open crystal structure databases, not relying on proprietary data sources, enabling it to adapt to the synthesis chemical space on-the-fly. In this manner, XERUS can suggest pre-refined multi/single phase results and provides an interface for optimizing one-shot Rietveld analysis. It was designed to be used either in Jupyter notebooks\cite{jupyter}, as an interactive tool with researchers, or to be adapted so it can be plugged-in directly into experimental workflows, providing a quickly analysis of the results. In the following sections, the core functions of XERUS and the method it uses to attempt phase identification are described. We also present analysis results for four different systems and we benchmark the results against the mixture dataset of Ref. \cite{szymanski2021probabilistic}.

\section{XERUS Description}
\begin{figure}[htb]
  \includegraphics[width=178mm]{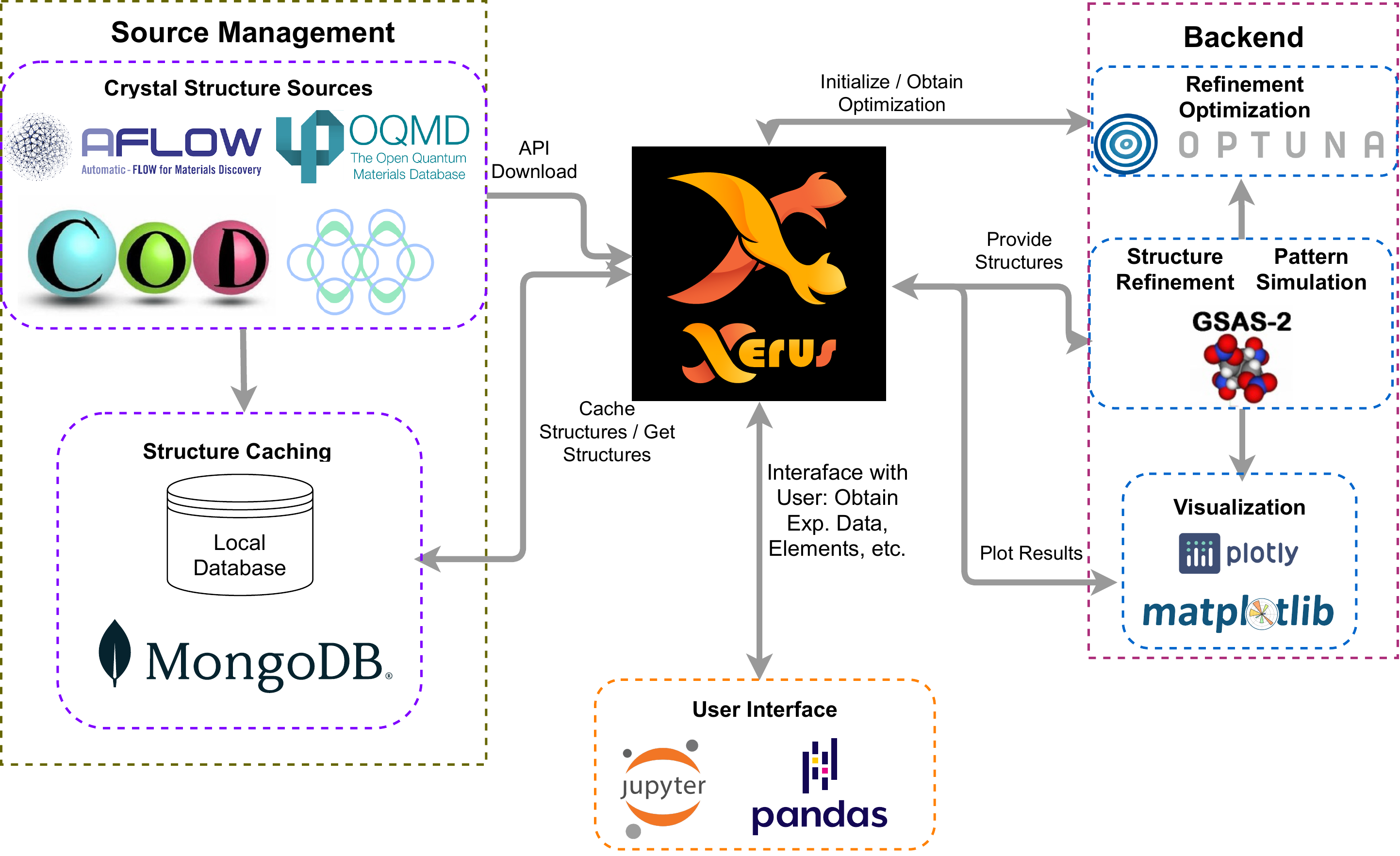}
  \caption{Overview of XERUS and its main building blocks, showing how crystal structure is obtained, how the patterns are simulated
  and how the data can be visualized.}
  \label{fig-overview}
\end{figure}

\begin{figure}[!htpb]
  \includegraphics[width=\linewidth]{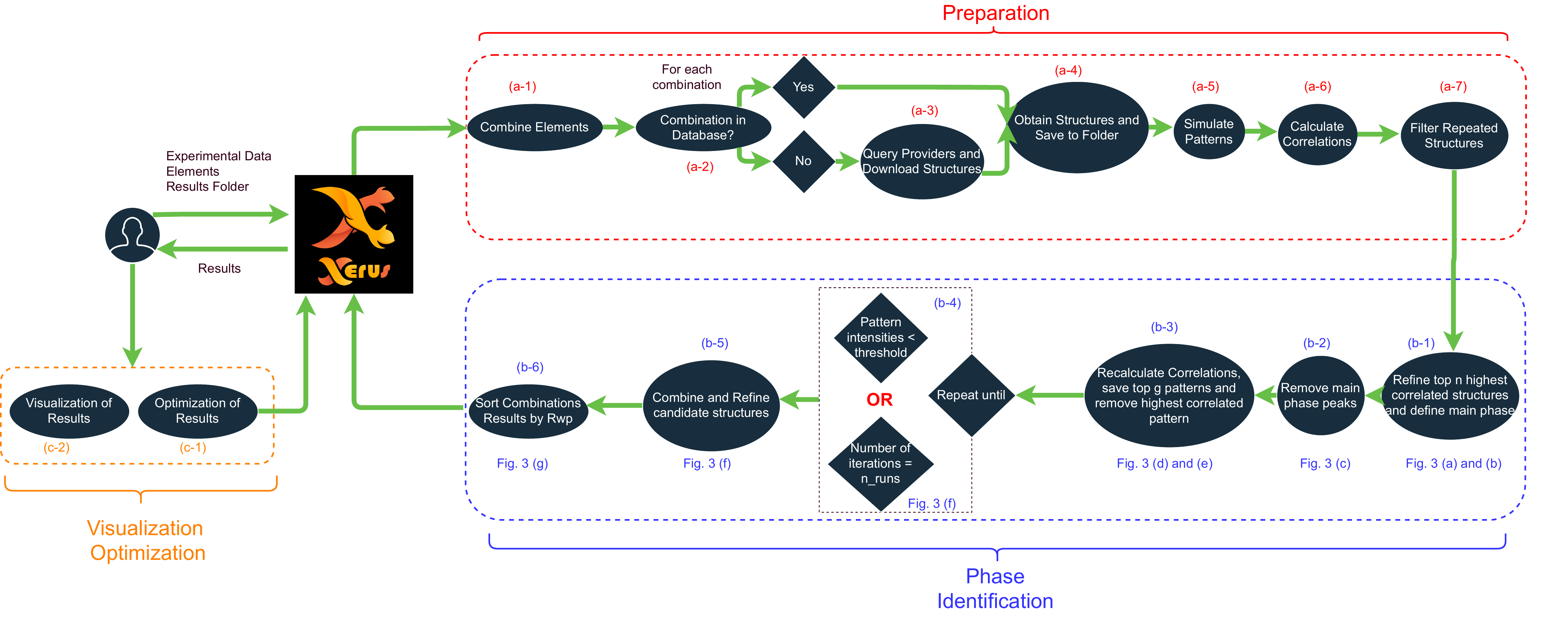}
  \caption{Workflow diagram of steps taken by XERUS after receiving the data with given parameters such as elements and directory
  in which to save. The preparation blocks present the steps from combining elements, to querying structures, calculating patterns and a first
  calculation of correlation, followed by structure filtering. This is then followed by the phase identification loop described in \textbf{Section} \ref{section-phase-id}
  and, finally, the results are reported to the user.}
  \label{fig-workflow}
\end{figure}

Below we describe the main three main building blocks of XERUS. 
\begin{itemize}
  \item Crystal Structure Management: Responsible for querying the MP, OQMD, AFLOW and the Crystallographic Open Database (COD), parsing and caching the structures for a given chemical space. 
  \item Backend: Responsible for pattern simulation, structure refinement, optimization and visualization. Provides the interface for phase identification.
  \item Interface: Allows for ease of visualizing and analyzing the obtained results, mostly through notebooks or by exporting the data.
\end{itemize}
These three building blocks are summarized in \textbf{Figure} \ref{fig-overview}. \textbf{Figure} \ref{fig-workflow} shows
the workflow of how XERUS works: from receiving the data, elements and directory to save the results as input, going through phase identification, and finally reporting the results to the user, while \textbf{Figure} \ref{fig-algo} shows all the steps done by XERUS in the phase identification process.  All these steps  will be discussed in the sections below.
\subsection{Preparation}
To easily adapt to any given chemical space, XERUS takes advantage of the computational databases API, such as AFLUX\cite{Rose2017}, the Materials Project API\cite{Ong2015} (MAPI), interfaced using pymatgen\cite{Ong2013}, the OQMD API and also the COD\cite{Graulis2009} REST API, to obtain the relevant crystal structures for the space. Here, XERUS also uses a local MongoDB\cite{mongodb} server, to locally cache the structures for future uses.\\
Then, for a given list of elements, (eg. Ho, B and O), XERUS creates the chemical space by first building all possible combinations of each element [\textbf{Figure} \ref{fig-workflow} (a-1)], in this case Ho, B, O, Ho-B, Ho-O, B-O and Ho-B-O. Afterwards, XERUS checks if each combination is already present in the local database [\textbf{Figure} \ref{fig-workflow}(a-2)]. If they are not found in the local database, XERUS will automatically query the structures of the  missing combination of the input elements from the databases and obtain the structures in CIF format [\textbf{Figure} \ref{fig-workflow} (a-3)], that will be used for simulation of patterns and refinement, described in more detail the in \textbf{Section} \ref{section-phase-id}. 
The acquired structures are then parsed by pymatgen to obtain metadata such as composition, space group, lattice system, and so on, and finally is they are saved into the the local database. After all structures are obtained, they are finally saved into the designed working directory [\textbf{Figure} \ref{fig-workflow} (a-4)].

\subsection{Phase Identification}\label{section-phase-id}
To achieve phase identification, XERUS implements a mixed approach of correlation (similarity) calculations with quick Rietveld refinements, followed by a pattern reduction, for the case of multiphase identification. Correlation metrics have been widely studied for use in XRD phase identification\cite{szymanski2021toward,long2007rapid,Iwasaki2017}, being able to provide an accurate construction of phase diagrams in combinatorial thin-film synthesis approach. Based on these previous results, here we use the Pearson correlation metric $C_{ij}$, where \textit{i} and \textit{j} are a given spectra with mean
value $\bar{i}$ and $\bar{j}$, is defined as\cite{long2007rapid}: 

\begin{equation}\label{correlation}
  C_{ij} = \sqrt{\frac{\sum_{k} (i_{k} - \bar{i})(j_{k} -\bar{j})}{\sum_{k} (i_{k} - \bar{i})^{2}\sum_{k} (j_{k} - \bar{j})^{2}}} 
\end{equation}
Then by simulating the patterns of the obtained structures for a given chemical space using GSASII\cite{Toby2013,o2018scripting} [\textbf{Figure} \ref{fig-workflow} (a-5)], the Pearson correlation between these patterns and the obtained experimental data can be quickly calculated using the pandas package\cite{reback2020pandas,mckinney-proc-scipy-2010} [\textbf{Figure} \ref{fig-workflow} (a-6)]. Note, that here we will always be calculating the correlations between simulation and the experimental data, therefore the correlation value between a simulated pattern \textit{k} and the experimental data will be noted as $C_{k}$. \\ 

\begin{figure}
  \includegraphics[width=120mm]{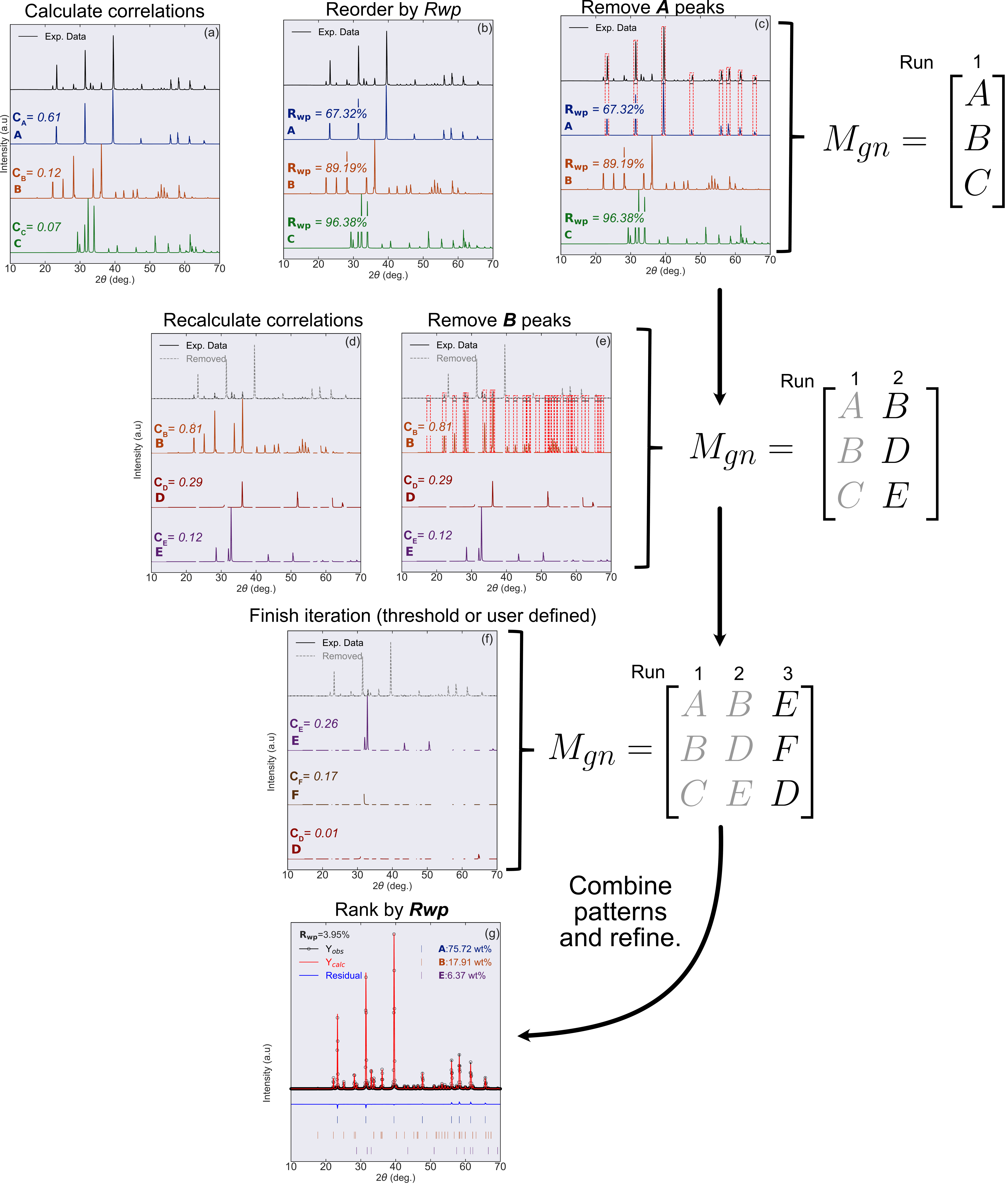}
  \caption{Phase detection approach of XERUS. (a): Top \textit{g} = 3 highest correlated patterns with the experimental data, 
  (b): Patterns re-ranked by $R_{wp}$ after a quick Rietveld refinement has been performed, 
  (c): The main phase is identified and the associated bragg position peaks are removed by a box of width $\delta$ degrees and 
  these patterns are saved in the refinement matrix $M_{gn}$, 
  (d): After the removal, the correlations are recalculated and the new \textit{g} patterns are selected and saved into the next column
  of the refinement matrix.
  (e): Now the highest correlated pattern is selected as the next possible phase,
   and again its associated bragg positions are removed from the data.
   (f): After a certain number of runs or the experimental data intensity has dropped below a threshold,
   XERUS stops this process. In the case of multiphase detection, these patterns are also considered as possible secondary phases.
   (g): The patterns in the refinement matrix are combined and refined and ranked by lowest $R_{wp}$. Here, we assume that
   the lowest $R_{wp}$ combination is the answer for the problem.}
  \label{fig-algo}
\end{figure}

\begin{figure}[!htbp]
  \includegraphics[width=178mm]{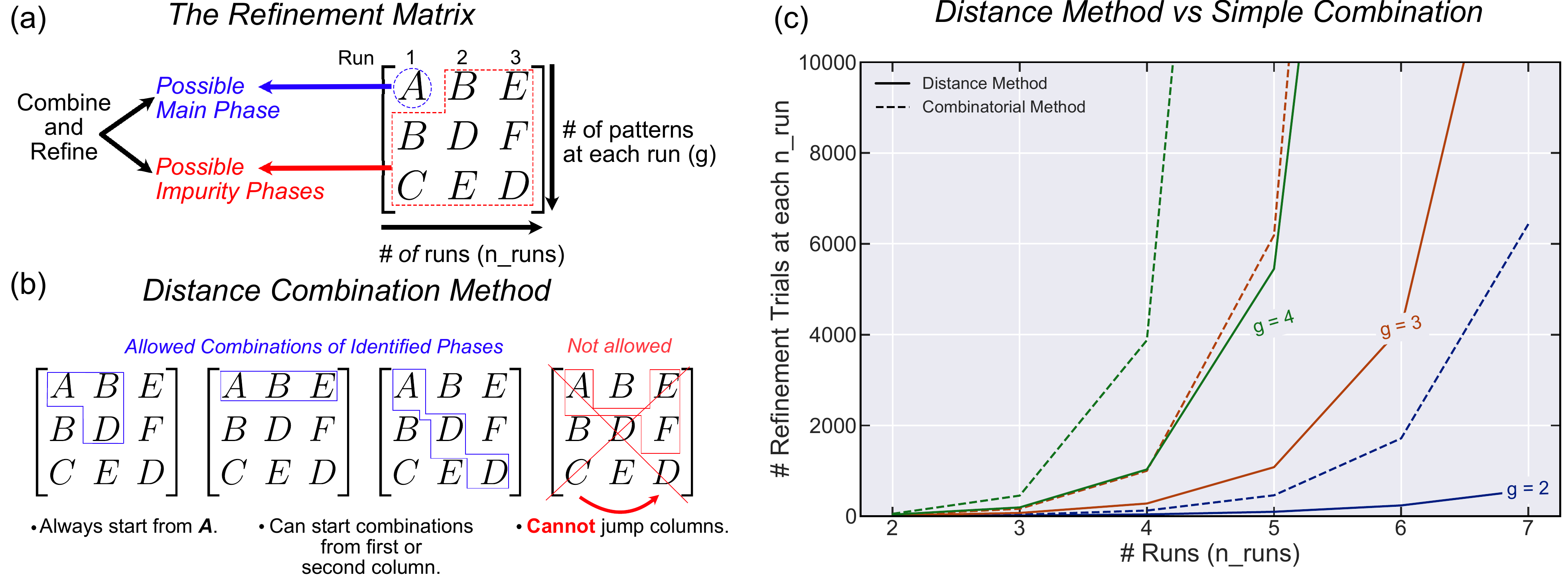}
  \caption{ (a) The refinement matrix definition. The possible main phase is highlighted by the blue circle,
  while the possible secondary phases are in red. (b) The overview of the distance-based approach for combination of the phases.
  (c) A comparison between this method (solid lines) and simple combination (dashed lines) of all possible phases for different values of \textit{g}.
  }
  \label{fig-ref-matrix}
\end{figure}

After the calculation of $C_{k}$ for each simulated pattern, XERUS filter repeated structures [\textbf{Figure} \ref{fig-workflow} (a-7)] by taking only one structure for the pair of a given combination and spacegroup, that has the highest correlation value. For example, for the structures in the 'Ho-B' space with spacegroup \textit{P}6/\textit{mmm}, only the structure from this group with highest $C_{k}$ will be used. This step is necessary, since XERUS will obtain crystal structure from many different sources, there will be possibly repeated crystal structures for the same compound and as well off-stoichiometric variations of the compound. Afterwards, the top \textit{g} patterns with highest $C_{k}$, where \textit{g} is a user-defined parameter (default \textit{g}= 3), are then refined against the experimental data using GSAS II. Next, they are re-ranked by the lowest weighted profile residual factor ($R_{wp}$) value, defined as\cite{toby_2006rwp}:
\begin{equation}\label{rwp}
  R_{wp} = \sqrt{ \frac{\sum_{i} w_{i}(y_{i}^{obs} - y_{i}^{calc})^{2}}{\sum_{i}w_{i} \left(y_{i}^{obs}\right)^{2}}} 
\end{equation}
Where $w_{i}= \frac{1}{\sigma(y^{obs}_{i})^{2}}$ and $\sigma(y^{obs}_{i})$ is the standard uncertainty\cite{toby_2006rwp}. This is an attempt
into taking into account artifacts that can lead to shift in peak positions such as lattice parameter changes and zero shift, which
are one of the weaknesses of a correlation only approach\cite{szymanski2021toward,Iwasaki2017}. Here, we note that we do not perform
an optimal refinement, rather we just refine background function with fixed set of parameters, zero-shift, lattice constants and instrument parameters,
which is often enough to give a reasonable first model for the data, in our experience.
Then we define the possible \textit{main} phase of material as the one who achieved the lowest $R_{wp}$ [\textbf{Figure} \ref{fig-workflow}(b-1)] and then these \textit{g} patterns are saved into the first column of a refinement matrix $M_{g1}$. After defining the main phase, the whole data (experimental and simulated) is reduced by removing boxes of width $\delta$, where $\delta$ is a hyperparameter with default value of 1.3 degree appropriate to our laboratory X-ray source, around the 2$\theta$ corresponding to each Bragg position of the identified phase [\textbf{Figure} \ref{fig-workflow}(b-2)]. This is followed by a recalculation of $C_{k}$ for these reduced patterns, and we again take the top \textit{g} patterns as the next possible phases [\textbf{Figure} \ref{fig-workflow}(b-3)]. Since the experimental data has been reduced, we do not perform any extra refinement in this step. Instead, we reduce the data around the Bragg positions of the highest correlated pattern in this run. These structures are also saved in the next column of the refinement matrix, that is $M_{g2}$. This continues until [\textbf{Figure} \ref{fig-workflow}(b-4)]:
\begin{itemize}
  \item The maximum intensity of the experimental pattern drops below a threshold (default 10\%), this case is defined as the \textit{auto} mode.
  \item A number of runs \textit{n\_runs} has been elapsed, defined by the user. 
\end{itemize}
This process is illustrated in \textbf{Figure} \ref{fig-algo}(a)-(f) for the case of an ideal three phase search using default threshold of 10\%. It is worth mentioning that, in the case of multiphase identification, the top \textit{g} structures of the last run are also saved into the matrix. \\ 
Now we turn attention to the refinement matrix $M_{gn}$. It`s purpose is to allow the combination of the identified main phase,
which by construction lies in the $M_{11}$ position, with the possible secondary phases, that is the remaining \textit{g}(\textit{n\_runs} - 1) phases in the matrix to be finally refined as exemplified in \textbf{Figure} \ref{fig-ref-matrix}(a).
To do this, ideally one could just then refine the identified main phase with all possible combinations of the secondary phases. This would entail, a total of $\frac{k!}{(k-w)!w!}$, where $k = g(n\_runs - 1)$ and $w = n\_runs - 1$, phases for refinement, leading to strong combinatorial explosion as \textit{g} and
\textit{n\_runs} increases. To soften this explosion, we define a distance-based method on how the patterns can be combined in the matrix:
\begin{itemize}
  \item We always start from $M_{11}$, that is phase \textbf{A} in the example matrix of \textbf{Figure} \ref{fig-algo}
  \item We allow the first pattern to start making combinations from the \textbf{first} or \textbf{second} column.
  \item Runs cannot be skipped. For instance, in \textbf{Figure} \ref{fig-ref-matrix}(b) case, the combination \textbf{A}-\textbf{E}-\textbf{F} is not allowed.
\end{itemize}
These rules are illustrated in \textbf{Figure} \ref{fig-ref-matrix}(b), and in \textbf{Figure} \ref{fig-ref-matrix}(c) we  show the comparison of the required number of refinements for a simple combinatorial method and the distance-based method. 
It is evident then, that the distance-based approach effectively suppresses the combinatorial explosion even if XERUS has to consider the presence of multiple phases, enabling their quick identification. Here, we stress the benefit of having several runs while using our similarity based approach. When a sample has multiple phases with minor fractions in addition to the main phase, the one with highest intensity, those minor phases usually will not have high similarity score in the first run, where the main phase is still present. To account for possible extra phases in the XRD pattern, XERUS reduces the experimental and simulated pattern, with the expectation that the similarity between the remaining XRD pattern might show higher score with the reduced simulated pattern for the minor phases in the sample, in the next runs. Indeed, it is quite often that the identified secondary phases, or minor phases, are ranked lower than \textit{g} in the first run, as it will be shown in the experimental results in \textbf{Section} \ref{results-hob2} (phase \textbf{D} that appears within \textit{g} = 3, only in the second run, and it was identified as the secondary phase) and in \textbf{Section} \ref{results-lsco} (phase \textbf{F} also only appears within \textit{g} at the second run). \\
Here, we also note the case of how XERUS deal with overestimation of number of phases, that is when the inputted \textit{n\_runs} is higher than the actual number of phases in the sample. For a given \textit{n\_runs}, XERUS considers all possible number of total present phases, including lower number of \textit{n\_runs}. Namely, suppose \textit{n\_runs} = 3, the $R_{wp}$ not only for three-phase refinements but also for single and two-phase are listed, so XERUS ranks all possible cases of phase combinations within a given \textit{n\_runs}. In this manner, XERUS avoid the overestimation of number of phases. Finally, all the considered phases and their $R_{wp}$ are listed in the result output, where the highest rank would be considered the the result in the automatic case, while also giving the option for a final human check, if needed.

\subsection{\label{section-opt}Optimization (Optional)} 
The main purpose of XERUS is to find out what are the secondary phases other than the objective material allowing a quick identification of the synthesis success. Depending on the percentage of such phases, adapting the synthesis method to minimize such phases, if possible, would be ideal. However, here we also acknowledge that phase determination is also not the only issue when dealing with XRD data. Often, it is desired to obtain structural information,
such as lattice parameters, atomic positions, phase fractions (in the case of multiphase patterns), and so on. This involves fitting
structural models such as Rietveld, which usually requires a initial guess of starting parameters followed
by sequential refinements where the parameters to be refined are varied\cite{oneshotriet-acs-2020}, to achieve optimal results. This is often done by a human expert, using a trial and error approach guided by their own experience, which can be time consuming. \\
To solve this issue, Ozaki \textit{et al.} proposed the use of an blackbox optimization approach (BBO)\cite{Ozaki2020-bbo}. In this setup, $R_{wp} (\mathbf{X})$ is treated as the function to optimize where $\mathbf{X}$ refers to a given input parameters fora refinement. \\
XERUS adopts their method, and adapts it to allow for multiphase refinement case. It can directly use obtained pre-refined patterns after the phase detection part, without the need of downloading or preparing
the search space, allowing for the obtained results to be quickly optimized [\textbf{Figure} \ref{fig-workflow} (c-1)] using the optuna package\cite{optuna_2019}. Here we note that, as discussed before, the quick Rietveld refinements used for phase identification are not optimal, and they are just used to help the phase identification process, and thus need to be optimized. For more details in this approach, we refer the readers to Ref. \cite{Ozaki2020-bbo}, where are in-depth discussion is provided.

\subsection{User Interface}
XERUS provides an low-code, easy-to-use API for exploring the results of phase identification and optimization (if done). All he results are saved into a pandas DataFrame, ordered by lowest $R_{wp}$ as discussed in \textbf{Section} \ref{section-phase-id}. It automatically parses GSASII files and provides out-of-the box static or interactive plots allowing for a quick inspection of the obtained results. On top of that, it can start the refinement optimization method described in \textbf{Section} \ref{section-opt} of any of the obtained combinations [\textbf{Figure} \ref{fig-workflow} (c-2)]. \\
XERUS also automatically exports almost all of the data used for phase identification such as phase identification results, pattern simulations, CIF structures, and optimization results, which can all be readily visualized with third-party software if necessary. \\
The usage of these methods are explained in the example notebook that will be available at \\
https://www.github.com/pedrobcst/Xerus/Examples/

\section{Results and Discussion}
To show how XERUS works, we will divide results into experimental data of multiphase materials into two sections. In the first section, we will describe the User-guided process where the \textit{n\_runs} (number of phases to search) is given as an input. Here we will show the role of keeping the human in-the-loop for the final decision on the results. In the second part, we will show the straight results from XERUS in automatic mode, where the final result is taken as the one that achieved the lowest $R_{wp}$ as discussed before. This will be done in the Li-Mn-Ti-O-F mixture dataset of Ref. \cite{szymanski2021probabilistic}, where in the end we will benchmark our results against
their deep learning approach trained specifically for this chemical space. Here we note that, for all cases, we used the default values of the box width $\delta$ = 1.3, and the number of patterns considered at each run, \textit{g} = 3. The use of these parameters, in our experience, is the one that give the best results in all the patterns we have tested, and these parameters might require adjustment depending on the data. \\
We also note that, the compositions listed in this section are not an attempt at identifying the synthesized material stoichiometry. Rather, they reflect the composition parsed for a given structure contained at the database of a given provider. They were also modified, when required, to the nominal composition of a material for clarity. For example, in the case of the parsed composition Mo$_{8}$C$_{4}$,we used Mo$_{2}$C instead. The as-parsed compositions are extensively listed in the Supplementary Information.
\subsection{User-guided process}\label{section-results}
To illustrate the use cases of XERUS, we prepared four different samples:
\begin{itemize}
  \item A simple mixture of Ni and Cr powders
  \item  An alloy of arc-melted Ho and B with 1:2.1 mol proportion, where the target material is HoB$_{2}$. The extra amount of boron is to lead to formation of a HoB$_{4}$ secondary phase.
  \item A sample of La$_{1-x}$Sr$_{x}$CuO$_{4}$ ($x$ = 0.25), as an example of an oxide material.
  \item An alloy of arc-melted Ho, Mo, C where the target compound is HoMoC$_{2}$.
\end{itemize}
\subsubsection{Ni-Cr powder mixture}
\begin{figure}[htpb]
  \includegraphics[width=\linewidth]{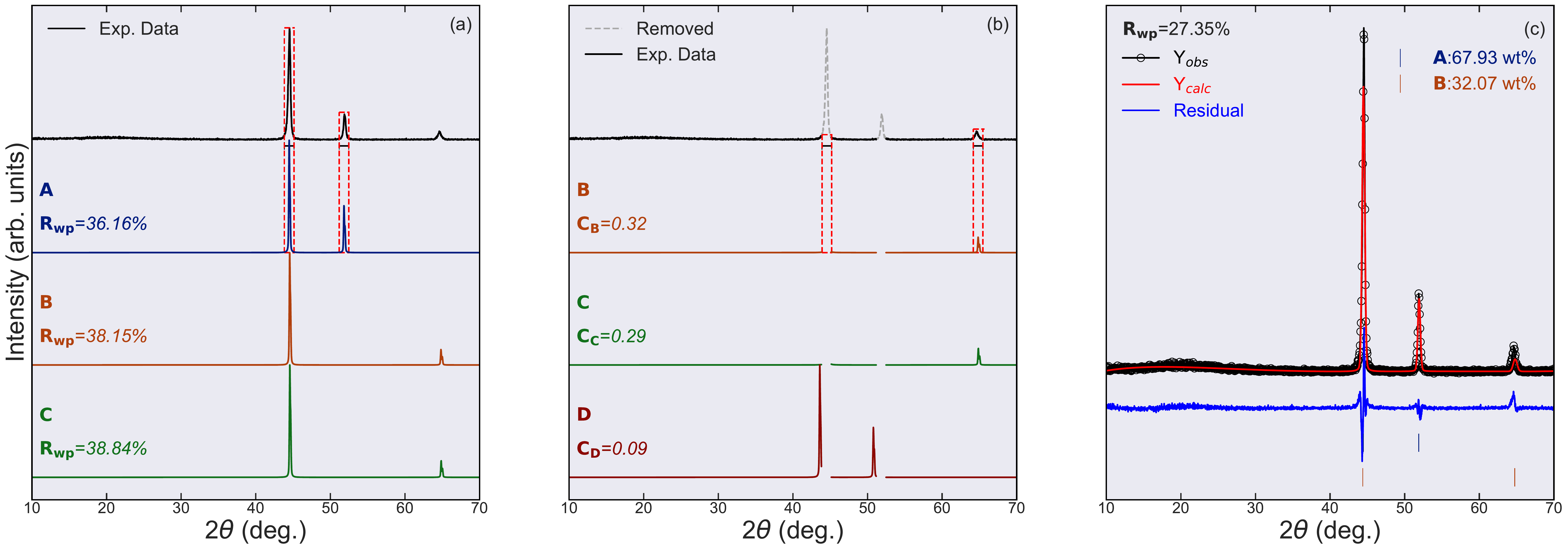}
  \caption{XERUS Results for a simple mixture of Cr-Ni. (a) Identification of main phase by $R_{wp}$ ranking. Here \textbf{A} refers to
  the cubic Ni \textit{Fm}$\bar{3}$\textit{m} phase, \textbf{B} is cubic \textit{Im}$\bar{3}$\textit{m} Cr phase and \textbf{C} is an alloy of cubic \textit{Im}$\bar{3}$\textit{m} Cr$_{0.8}$Ni$_{0.2}$.(b) Second phase identification after the removing the peaks of \textbf{A}.
   \textbf{D} refers to the theoretical cubic \textit{Fm}$\bar{3}$\textit{m} Cr phase. (c) Final result after refinement of all combinations between identified phases where the final shows the mixture of cubic Cr and Ni as expected.}
  \label{fig-crni}
\end{figure}

The Ni and Cr powder mixture represents the easiest case of two-phase mixture identification, 
since both Ni and Cr have non-overlapping peaks, and serves as an example where the final answer is clearly known, since no reaction of the elements takes place. \\
\textbf{Figure} \ref{fig-crni} (a)-(b) shows the XERUS process applied for this mixture with \textit{n\_runs} = 2. The phase cubic \textit{Fm}$\bar{3}$\textit{m} Ni phase (\textbf{A}) was identified as the possible main phase, with the possible secondary phases: cubic \textit{Im}$\bar{3}$\textit{m} Cr (\textbf{B}), the Cr-Ni Cr$_{0.8}$Ni$_{0.2}$ alloy (\textbf{C}) with same crystal structure as Cr, and the cubic \textit{Fm}$\bar{3}$\textit{m} theoretical Cr structure\cite{Cr-theo-Fm3m-PRB-1993} (\textbf{D}). \\
Here we note that all the secondary phases have the same structure as either Ni or Cr, with exception of the theoretical Cr phase,and thus have the same pattern as the pure elements. \textbf{Figure} \ref{fig-crni}(c)
shows the final combination chosen as the answer: cubic Ni and cubic Cr as expected, which achieved the second lowest $R_{wp}$.
The lowest $R_{wp}$ combination was between Ni and the Cr-Ni alloy. Since this alloy has the same structure as pure Cr, this
combination would also be acceptable as the answer. We note that the difference between those two combinations in the $R_{wp}$ ranking was less than 1\% (See \textbf{Table} SM. 1). 

\subsubsection{\label{results-hob2}Ho-B alloy}
To represent a more realistic experimental situation of a binary system,
we prepared a Ho-B alloy by arc melting. The Ho-B phase diagram\cite{wei2020-ho-b-phasediagram} is simple 
and provides a straightforward answer to which phases can exist. 
To attempt to obtain a simple two-phase system, the target material was HoB$_{2.1}$, which should lead to a mixture of both HoB$_{2}$ and HoB$_{4}$ phases. \\ 
\begin{figure}[!htb]
  \includegraphics[width=\linewidth]{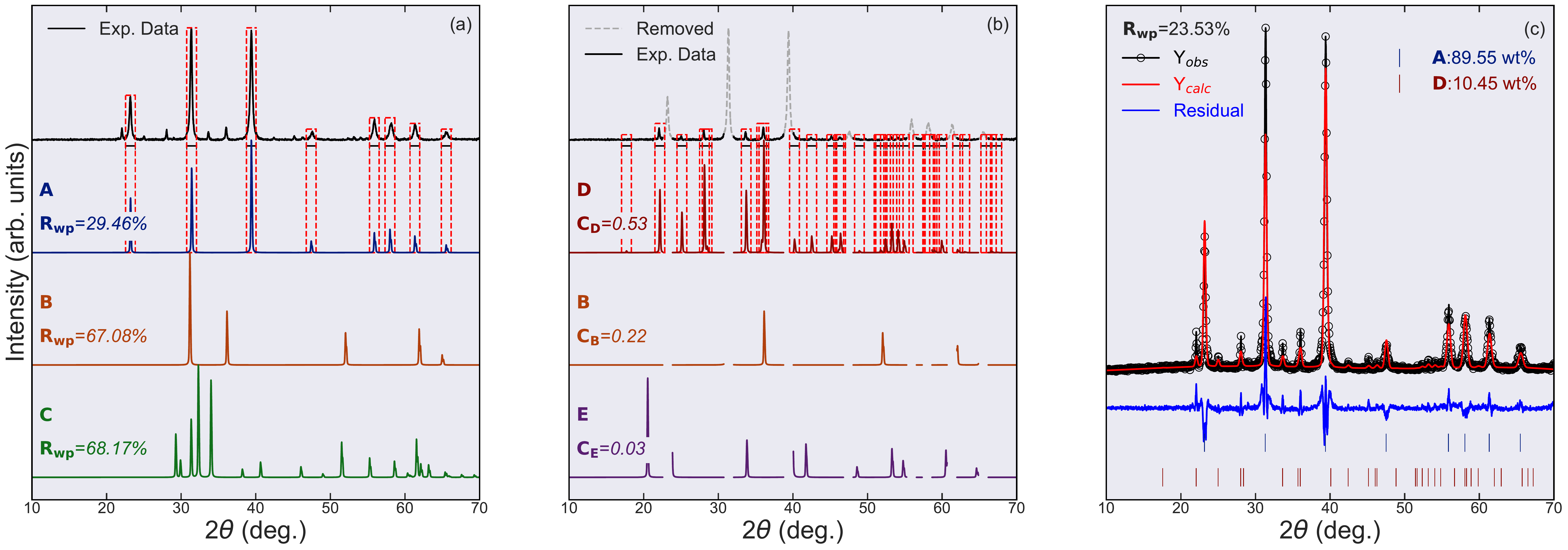}
  \caption{XERUS Results for the obtained alloy of Ho-B. 
  (a) Identification of main phase by $R_{wp}$ ranking. \textbf{A}: hexagonal \textit{P}6/\textit{mmm} HoB$_{2}$ , \textbf{B}: cubic \textit{Fm}$\bar{3}$\textit{m} Ho , \textbf{C}: trigonal \textit{R}$\bar{3}$\textit{m} Ho. (b) Second run of XERUS process. \textbf{D}: tetragonal \textit{P}4/\textit{mbm} HoB$_{4}$ and \textbf{E}: cubic \textit{Fm}$\bar{3}$\textit{m} HoB$_{12}$. (c) Final result after refinement of all combinations between \textbf{A} and the possible secondary phases, showing that the obtained sample is a mixture of HoB$_{2}$ and HoB$_{4}$.}
  \label{fig-hob}
\end{figure}
\textbf{Figure} \ref{fig-hob} shows the XERUS process for this sample with \textit{n\_runs} = 2. \textbf{Figure} \ref{fig-hob}(a) shows the first phase identification done by XERUS, where the main phase is found as a hexagonal \textit{P}6/\textit{mmm} HoB$_{2}$ (\textbf{A}). The other structures belong to different metastable structures of Ho: cubic \textit{Fm}$\bar{3}$\textit{m} Ho (\textbf{B}) and trigonal \textit{R}$\bar{3}$\textit{m} Ho (\textbf{C}). 
\textbf{Figure} \ref{fig-hob}(b) shows the new patterns after the peaks belonging to the HoB$_{2}$ phase have been removed. In this case, two new possible phases appear: tetragonal \textit{P}4/\textit{mbm} HoB$_{4}$ (\textbf{D}), with the highest correlation value, and cubic \textit{Fm}$\bar{3}$\textit{m} HoB$_{12}$ (\textbf{E}). The final result obtained by XERUS is shown in \textbf{Figure} \ref{fig-hob}(c), showing that the sample is indeed a mixture of HoB$_{2}$ and HoB$_{4}$, which is also the combination that achieved the lowest $R_{wp}$ (See \textbf{Table} SM2).

\subsubsection{\label{results-lsco}LSCO}
\begin{figure}[htpb]
  \includegraphics[width=\linewidth]{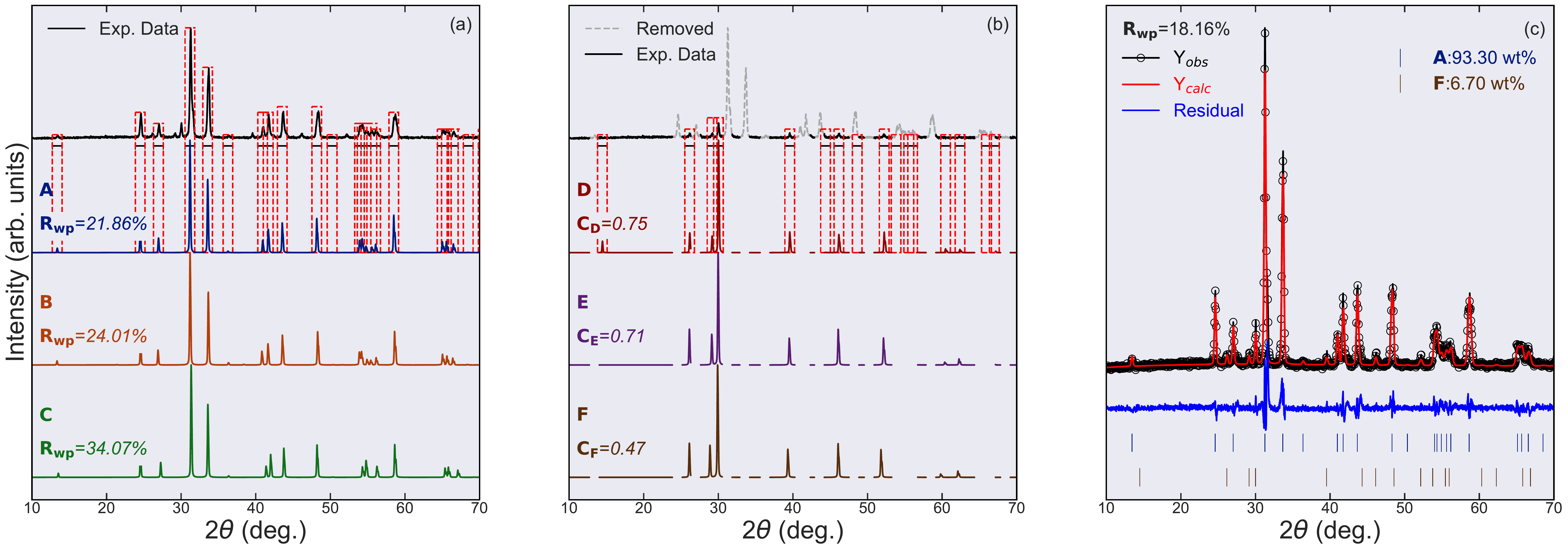}
  \caption{XERUS Results for the obtained LSCO sample. 
  (a) Identification of main phase by $R_{wp}$ ranking. \textbf{A}: tetragonal \textit{I}4/\textit{mmm} LSCO (La$_{1.85}$Sr$_{0.15}$CuO$_{3.6}$), \textbf{B}: orthorhombic \textit{Cmce} LSCO (La$_{1.87}$Sr$_{0.13}$CuO$_{4}$)  and \textbf{C}: tetragonal \textit{I}4/\textit{mmm} La$_{2}$CuO$_{4}$. (b) Second run of XERUS process. \textbf{D}: trigonal \textit{P}$\bar{3}$21 La$_{2}$O$_{3}$, \textbf{E}: hexagonal \textit{P}6$_{3}$/\textit{mmc} La$_{2}$O$_{3}$ and \textbf{F}: trigonal \textit{P}$\bar{3}$\textit{m}1 La$_{2}$O$_{3}$. 
  (c) Final result after refinement of all combinations between \textbf{A} and the possible secondary phases,
  showing that the obtained sample is a mixture of tetragonal LSCO and trigonal La$_{2}$O$_{3}$.}
  \label{fig-lsco}
\end{figure}

As an illustration of the application to an oxide ceramic material, we apply XERUS with \textit{n\_runs} = 2, to a sample of polycrystalline LSCO synthesized by solid-state reaction, with the larger chemical space of Sr-La-Cu-O. \\
 \textbf{Figure} \ref{fig-lsco}(a) shows the result for the main phase search. It is identified as the tetragonal \textit{I}4/\textit{mmm} LSCO structure with composition La$_{1.76}$Sr$_{0.15}$CuO$_{4}$ (\textbf{A}). Here we also note that other phases are also both related to LSCO: orthorhombic \textit{Cmce} LSCO La$_{1.87}$Sr$_{0.13}$CuO$_{4}$ (\textbf{B}) and the undoped tetragonal \textit{I}4/\textit{mmm} La$_{2}$CuO$_{4}$. (\textbf{C}). The orthorhombic LSCO structure probably corresponds to the low-temperature, as this material is known to undergo a tetragonal-orthorhombic structural transition at lower temperatures\cite{lsco-day1987temperature,lsco1-radaelli1994structural}. 
\textbf{Figure} \ref{fig-lsco}(b) shows clearly that the sample has at least a secondary phase. Interestingly, they are all suggested to be La$_{2}$O$_{3}$, one of the starting materials used for sample synthesis: the highest correlated pattern \textbf{D} trigonal \textit{P}$\bar{3}$21 La$_{2}$O$_{3}$ (\textbf{D}), hexagonal \textit{P}6$_{3}$/\textit{mmc} La$_{2}$O$_{3}$ (\textbf{E}) and trigonal \textit{P}$\bar{3}$\textit{m}1 La$_{2}$O$_{3}$ (\textbf{F}). Note that all these structures possess extremely similar patterns. Indeed their existence is due to diverging literature on the correct structure of La$_{2}$O$_{3}$ where diverse authors reported possible structures for this material\cite{la2o3-structure} and thus they all exist at the COD database. 
\textbf{Figure} \ref{fig-lsco}(c) shows the final result of the combination with lowest obtained $R_{wp}$ (See \textbf{Table} SM3.). It shows a combination of \textit{I}4/\textit{mmm} tetragonal LSCO as the main phase and \textit{P}$\bar{3}$\textit{m}1 trigonal La$_{2}$O$_{3}$ with a large wt\%, an indication that not all the starting materials have reacted. We note that in this situation, since all the patterns of LSCO and La$_{2}$O$_{3}$ are almost identical, XERUS could possibly find  the answer as a combination of any of these structures. Such a case highlights the importance of a final human confirmation for an accurate determination of the phases, that is depicted in \textbf{Figure} \ref{fig-lsco}(c).

\subsubsection{Ho-Mo-C alloy}
\begin{figure}[htpb]
  \includegraphics[width=\linewidth]{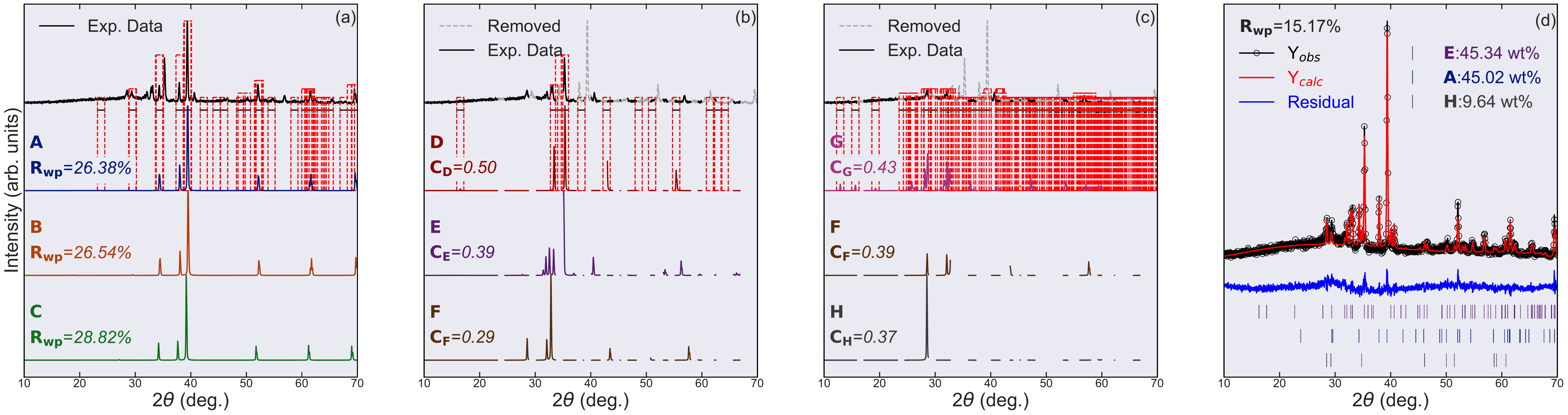}
  \caption{XERUS Results for the obtained alloy of Ho-B. 
(a) Identification of main phase by $R_{wp}$ ranking. \textbf{A}: orthorhombic \textit{Pbcn} $\alpha$-Mo$_{2}$C, 
\textbf{B}: orthorhombic \textit{Pca}2$_{1}$ $\beta$-Mo$_{2}$C, \textbf{C}: trigonal \textit{P}$\bar{3}$1\textit{m} Mo$_{2}$C.
(b) Second run of XERUS process. \textbf{D}: hexagonal \textit{P}6$_{3}$/\textit{mmc} MoC, \textbf{E}: orthorhombic \textit{Pnma} HoMoC$_{2}$, and \textbf{F}: hexagonal \textit{P}6$_{3}$/\textit{mmc} Ho.
(c) Third run search. \textbf{G}: monoclinic \textit{P}2$_{1}$/\textit{c} Ho$_{4}$C$_{7}$ and \textbf{H}:  tetragonal HoC$_{2}$ \textit{I}4/\textit{mmm}
(d) Final result after refinement of all combinations, including two phase mixture and three phase mixture
 between \textbf{A} and the possible secondary phases. The final result shows that the sample is possibly
 a combination of $\alpha$-Mo$_{2}$C, HoMoC$_{2}$ and HoC$_{2}$.}
\label{fig-homoc2}
\end{figure}

As an example of a possible data-driven material, we attempted to synthesize a sample of HoMoC$_{2}$. \\
HoMoC$_{2}$ is a material whose physical properties have never been reported. Only its crystal structure is known\cite{homoco2-jeitschko1986ternary}, present in the MP database but not in the COD. Moreover, the Ho-Mo-C phase diagram,
to the best of our knowledge, has never been reported in the literature, making the prediction of the possible secondary phases that can appear in this system not straightforward.
\textbf{Figure} \ref{fig-homoc2} shows the XERUS process applied to this alloy with \textit{n\_runs} = 3. \textbf{Figure} \ref{fig-homoc2}(a) shows the main phase identification of this alloy. Interestingly, it was identified as the orthorhombic \textit{Pbcn} $\alpha$-Mo$_{2}$C (\textbf{A}). The other two proposed phases are the orthorhombic \textit{Pca}2$_{1}$ $\beta$-Mo$_{2}$C (\textbf{B}) and high temperature trigonal \textit{P}$\bar{3}$1\textit{m} Mo$_{2}$C phase\cite{Mo2C-high-T}. In the next run, depicted in \textbf{Figure} \ref{fig-homoc2}(b), the highest correlated pattern is found to be \textit{P}6$_{3}$/\textit{mmc} MoC (\textbf{D}), followed by the target material orthorhombic \textit{Pnma} HoMoC$_{2}$ (\textbf{E}), and the hexagonal \textit{P}6$_{3}$/\textit{mmc} Ho (\textbf{F}). In the final run, shown in \textbf{Figure} \ref{fig-homoc2}(c), two new candidates phases are found, with the highested correlated with the remaining peaks being the monoclinic \textit{P}2$_{1}$/\textit{c} Ho$_{4}$C$_{7}$ (\textbf{G}), and the last one being the  \textit{I}4/\textit{mmm} HoC$_{2}$ (\textbf{H})
The final result, suggested by XERUS is illustrated in \textbf{Figure} \ref{fig-homoc2}(d), where the lowest $R_{wp}$ combination shows that this alloy consists of at least of a mixture of $\alpha$-Mo$_{2}$C, HoMoC$_{2}$ and HoC$_{2}$ (See \textbf{Table} SM.4), indicating that the synthesis of the target material was not successful. This case highlights the importance of the hyperparameter \textit{g}: both HoMoC$_{2}$ and HoC$_{2}$ were not the highest correlated phase but they lied within the top \textit{g}. Thus, if \textit{g} was set to a lower bound, such as 2, XERUS would miss the HoC$_{2}$ phase. Also, we stress here that such impurity phase identification and estimation of the phases ratios have a massive importance for optimizing the synthesis process/conditions.

\subsection{Automatic Process and Benchmarking}
To evaluate better the accuracy of XERUS, a comparison between it and the previously mentioned DL methods would be interesting. However, training the models described by previous works would be extremely time consuming and in most of the cases, they would need to be adapted to accept structures beyond the ICSD. In addition, most of the experimental dataset of these previous works are not openly available, with few exceptions. Therefore, here we benchmark XERUS against the Li-Mn-Ti-O-F experimental mixture dataset of Ref. \cite{szymanski2021probabilistic} for the case of multiphase classification, and we compare the results obtained with their optimized DL approach trained only for this chemical space. \\
This dataset consists of 10 different mixtures of Li$_{2}$TiO$_{3}$, MnF$_{2}$, LiMn$_{2}$O$_{4}$, Li$_{2}$MnO$_{3}$, MnO, TiO$_{2}$, LiMnO$_{2}$, LiF and Mn$_{2}$O$_{3}$, with five three-phase and five twophases mixtures, summarized in the first column of \textbf{Table} \ref{table-auto}. \\
To compare with the DL method in this dataset, we ran XERUS in using \textit{n\_runs} = \textit{auto} with the default values of $\delta = 1.3$, \textit{g} = 3. In this case, a threshold for the drop of the intensity maximum of the experimental data is necessary. Here, we use a threshold of 10\%, the default value of XERUS, in same manner as Ref. \cite{szymanski2021probabilistic}. \\

\begin{figure}[htpb]
  \includegraphics[width=\linewidth]{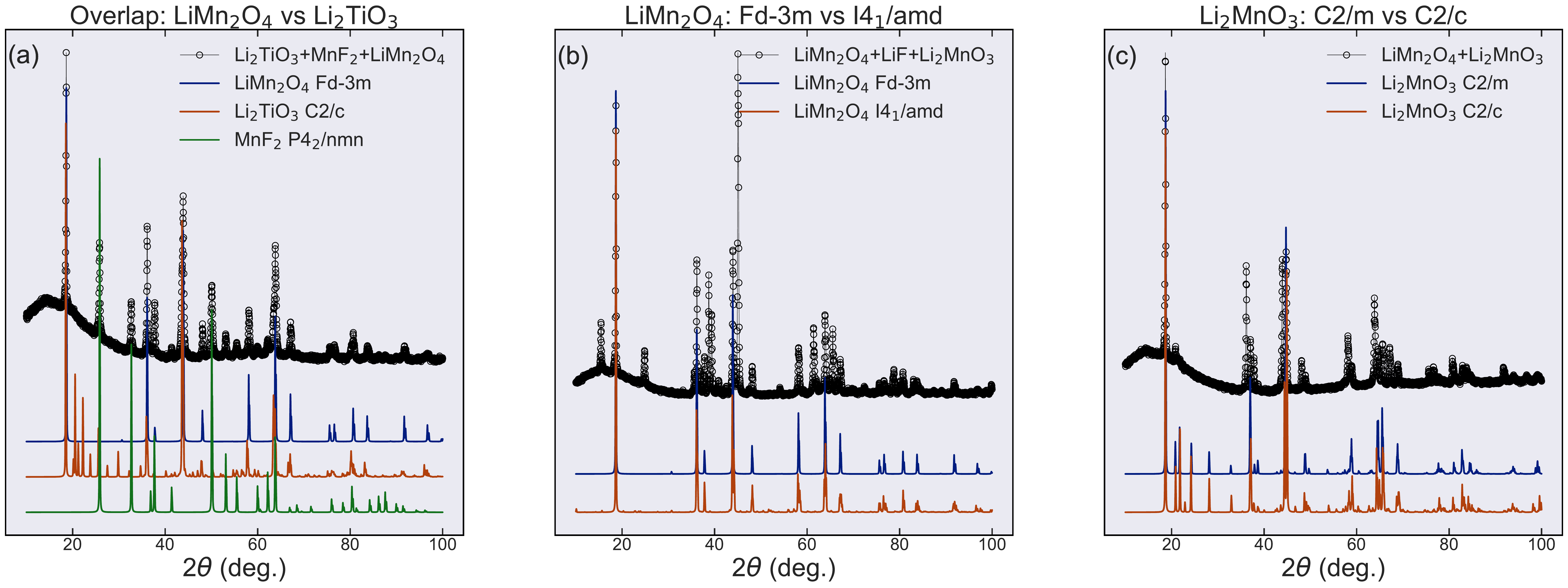}
  \caption{Identified pitfalls of XERUS. The black line with open dots refers to the experimental data while the solid lines are the patterns automatically simulated by XERUS. (a): A strong overlap between the peaks of the phases can lead XERUS to fail to find one or more of the highly overlapping phases. (b) and (c): Chemical compositions with different structures. (b) shows the room temperature and low-temperature structure of LiMn$_{2}$O$_{4}$ and (c) shows diverging structures reported in the literature regarding the crystal structure of Li$_{2}$MnO$_{3}$.}
\label{fig-overlap}
\end{figure}

The results obtained by XERUS are summarized in the second and third columns of \textbf{Table} \ref{table-auto}. Out of 25 possible phases, XERUS identifies correctly 22 phases with 2 classification mistakes and one phase not found, achieving an accuracy of 88\% in autonomous mode, a result that is comparable to the DL approach that scored a 92\% accuracy for multiphase classification. To better understand the possible weaknesses of XERUS an investigation on the source of the errors in such a controlled dataset is important. \\
First, the inability of XERUS of finding one of the phases in the Li$_{2}$TiO$_{3}$+MnF$_{2}$+LiMn$_{2}$O$_{4}$ mixture highlights one of its weaknesses: highly overlapping patterns. \textbf{Figure} \ref{fig-overlap}(a) illustrates this case, where the simulated patterns of each material in the mixture and the experimental data is shown. Clearly there is a strong overlap between highest intensity peaks of Li$_{2}$TiO$_{3}$ and the other two structures. In such a case, since XERUS will reduce the patterns based on the Bragg positions of the \textit{"best"} candidate of a given run, thus the highest intensity peaks of the possible secondary phase will all be removed and thus XERUS will fail to detect them. In this particular case, LiMn$_{2}$O$_{4}$ is found as the \textit{main} possible phase and the first secondary phase as MnF$_{2}$. After the second run, the intensity drops below the threshold and the iteration stops and only two phases out of three is found. We note that, even if XERUS is ran specifying the number of iterations, in this case \textit{n\_runs} = 3, Li$_{2}$TiO$_{3}$ is not detected. \\
Now lets turn our attention to the two misclassifications. The first misclassification, where XERUS finds the tetragonal \textit{I}4$_{1}$/\textit{amd} LiMn$_{2}$O$_{4}$ phase is similar to what was observed for the case of LSCO: The LiMn$_{2}$O$_{4}$ crystallizes into a cubic \textit{Fd}$\bar{3}$\textit{m} structure at room temperature and upon cooling its structure changes. Early reports indicated that the cubic structure is distorted into a tetragonal \textit{I}4$_{1}$/\textit{amd} one\cite{limn2o4-wills1999low-tetragonal}. However later single crystal studies revealed that the structure is actually  orthorhombic\cite{limn2o4-single-orthorhombic}. Nethertheless, similar to LSCO, both these structures exists in the COD. In particular, the tetragonal distortion pattern is extremely similar to the room temperature one, as shown in \textbf{Figure} \ref{fig-overlap}(b), with exception of few low intensity peaks, that will probably be lost in the background. In this situation, it is natural for XERUS to suggest both phases as a solution. Indeed, the second best solution found by XERUS is with the cubic structure and the $R_{wp}$ difference between these solutions is as low as 0.02\% (See \textbf{Table} SM. 5). \\
The last misclassification is similar to the case described for La$_{2}$O$_{3}$: The structure of LiMn$_{2}$O$_{3}$ was first reported to belong to the \textit{C}2/\textit{c} space group. However, later it was confirmed to actually to have a \textit{C}2/\textit{m} symmetry\cite{li2mno3-boulin}. Similar to La$_{2}$O$_{3}$, both of these structures are present in the COD and thus are considered by XERUS. \textbf{Figure} \ref{fig-overlap}(c) shows the simulated pattern of both structures. Similar to the other cases, they are almost identical and the second best combination is indeed with the \textit{C}2/\textit{m} structure, with a $R_{wp}$ difference of 0.3\% between them (See \textbf{Table} SM. 6).  \\
Here we note an important difference between the inability to find one othe phases due to overlap and the two misclassifications due to similar structures: In the case of overlap, the correct phase is never found even when run manually with a given number of phases to search. In the other cases, the correct answer usually lies in the top 3 suggested combinations, which can be quickly checked in the end by a human. 
In this manner, we identify the major weaknesses of XERUS as situations where there is strong phase overlap. In the case of similar structures due to low-temperature phases and conflicting literature data, all the phases can be easily distinguished by a researcher with previous knowledge, and in this case XERUS only fails to identify one out of 25 phases, leading to 96\% accuracy. 
Therefore, XERUS can be exploited by researchers to obtain a quick identification of their synthesis results, allowing them to optimize the design of their synthesis method and conditions, possibly accelerating materials development.
\begin{table}
  \caption{Results of XERUS using automatic mode on the experimental data set of Ref. \cite{szymanski2021probabilistic}.The identified patterns are shown together with their space group. Here we take the result of XERUS as the combination that achieved the lowest $R_{wp}$. To classify as correct or incorrect, we take the answer as the stable structure at room temperature. Even if a structure with matching composition is found, if the structure found is a modification of the room temperature structure, for example low temperature, it is deemed incorrect. Note$^{*}$: The original mixture is described in the original data set as LiMn$_{2}$O$_{4}$+Li$_{2}$MnO$_{3}$+LiF. XERUS analysis indicates that this mixture might has been mislabelled, and LiMnO$_{2}$ was mixed instead of Li$_{2}$MnO$_{3}$.}
  \begin{ruledtabular}
  \begin{tabular}{@{}lll@{}}
    \hline
    Mixture & Correctly Identified & Mistaken [Not Found] \\
    \hline
    Li$_{2}$TiO$_{3}$+MnF$_{2}$+LiMn$_{2}$O$_{4}$ & LiMn$_{2}$O$_{4}$(\textit{Fd}$\bar{3}$\textit{m})+MnF$_{2}$(\textit{P}4${_2}$/\textit{mnm}) & [Li$_{2}$TiO$_{3}$ (\textit{C}2/\textit{c})]  \\
    Li$_{2}$MnO$_{3}$+MnO+TiO$_{2}$  & MnO(\textit{Fm}$\bar{3}$\textit{m})+Li$_{2}$MnO$_{3}$(\textit{C}2/\textit{m})+TiO$_{2}$(\textit{I}4$_{1}$/\textit{amd}) & - \\
    Mn$_{2}$O$_{3}$+TiO$_{2}$+LiMnO$_{2}$ & Mn$_{2}$O$_{3}$(\textit{Ia}$\bar{3}$)+LiMnO$_{2}$(\textit{Pmmn})+TiO$_{2}$(\textit{I}4$_{1}$/\textit{amd}) & -\\
    LiMn$_{2}$O$_{4}$+LiMnO$_{2}$*+LiF & LiF(\textit{Fm}$\bar{3}$\textit{m})+LiMnO$_{2}$(\textit{Pmmn}) & LiMn$_{2}$O$_{4}$(\textit{I}4$_{1}$/\textit{amd}) \\
    Mn$_{2}$O$_{3}$+Li$_{2}$TiO$_{3}$ & Mn$_{2}$O$_{3}$(\textit{Ia}$\bar{3}$)+Li$_{2}$TiO$_{3}$(\textit{C}2/\textit{c}) & - \\
    Mn$_{2}$O$_{3}$+LiF & Mn$_{2}$O$_{3}$(\textit{Ia}$\bar{3}$)+LiF(\textit{Fm}$\bar{3}$\textit{m}) & -  \\
    TiO$_{2}$+MnO & MnO(\textit{Fm}$\bar{3}$\textit{m})+TiO$_{2}$(\textit{I}4$_{1}$/\textit{amd}) & -  \\
    MnF$_{2}$+TiO$_{2}$ & MnF$_{2}$(\textit{P}4$_{2}$/\textit{mnm})+TiO$_{2}$(\textit{I}4$_{1}$/\textit{amd}) & - \\
    LiMn$_{2}$O$_{4}$+Li$_{2}$MnO$_{3}$ & LiMn$_{2}$O$_{4}$(\textit{Fd}$\bar{3}$\textit{m}) & Li$_{2}$MnO$_{3}$ (\textit{C}2/\textit{c}) \\
    LiMnO$_{2}$+Li$_{2}$MnO$_{3}$ & Li$_{2}$MnO$_{3}$(\textit{C}2/\textit{m})+LiMnO$_{2}$(\textit{Pmmn}) & - \\
    \hline
  \end{tabular}
\end{ruledtabular}
\label{table-auto}
\end{table}
\section{Conclusion}
In this work, we present the open source XERUS framework for semi-automatic/automatic analysis of XRD patterns. By taking advantage of materials simulations databases together with the COD, it can adapt on-the-fly to any chemical space, obtaining the relevant crystal structures. We show that the correlation method for pattern matching can be extended to multiphase classification, by employing a simple iterative peak removal process coupled with correlations calculations, where finally the found phases can be combined and refined using GSAS II. \\
We apply this method first to four different synthesized samples, ranging from a simple mixture to a complex alloy and show that XERUS can possibly identify all the phases in these materials. We also benchmark XERUS against a Li-Mn-O-Ti-F mixture dataset, and we show that XERUS running in automatic mode achieves an accuracy almost as good as previous deep-learning methods in the same dataset (88\% vs 92\%), without the need of simulation of large synthetic datasets,training of large models or use of proprietary databases. With these results, we also discuss the possible pitfalls of XERUS. \\
A tutorial with how to use XERUS, with the steps to reproduce all the results obtained in this paper will be available at https://www.github.com/pedrobcst/Xerus
\section{Experimental Section}\label{exp-sec}

\subsection{Sample Synthesis and Preparation}
La$_{2-x}$Sr$_x$Cu$O_4$(LSCO) with $x=0.25$ was prepared by solid-state reaction. The starting materials were La$_2$O$_3$, SrCO$_3$(99\%) and CuO($\geq99.9\%)$ powders. The La$_2$O$_3$ powder was prepared by dehydration of La[OH]$_3$ in a furnace at 1000°C overnight.
The powders were mixed and made into pellets with 10 mm. of diameter, afterwards placed on an alumina crucible and sintered in a furnace for 8 hours at 950°C and then cooled slowly. The pellets were made into powder and pelletized a second time, annealed again at 1000°C for 8 hours and cooled slowly.
The second sample, CrNi, was made by mixing commercial Cr and Ni powders. The third sample, HoB$_2$, was arc-melted from the elements in an Ar atmosphere using Ho and B metals with a stoichiometry of 1:2.1. The fourth sample, HoMoC$_2$, was also arc-melted in an Ar atmosphere with stoichiometry amounts Ho, Mo and C as starting materials. \\

\subsection{X-ray analysis}
Powder X-ray diffraction (XRD) analysis was done on a Rigaku MiniFlex 600 diffractometer using Cu-K$\alpha$ radiation, with 2$\theta$ ranging from 10° to 70° and a 2$\theta$ scanning step of 0.02°. 


\section{Acknowledgements}
We are grateful for fruitful discussions with Luca Foppiano. This work was supported by the JST-Mirai Program (Grant No. JPMJMI18A3), JSPS KAKENHI (Grant Nos. 19H02177, 20K05070, 20H05644, 20K22420), JSPS Bilateral Program (Grant No. JPJSBP120214602) and JST CREST (Grant No. JPMJCR20Q4). P.B. Castro acknowledges the scholarship support from the Ministry of Education, Culture, Sports, Science and Technology (MEXT), Japan.

\bibliography{references}

\end{document}



\title{Supplementary Information for \\ XERUS: \\ An open-source tool for quick XRD phase identification and refinement automation}

\author{Pedro Baptista de Castro}
\email{CASTRO.Pedro@nims.go.jp}
\affiliation{\NIMS}
\affiliation{\UT}
\author{Kensei Terashima}
\email{TERASHIMA.Kensei@nims.go.jp}
\affiliation{\NIMS}
\author{Miren Garbine Esparza Echevarria}
\affiliation{\NIMS}
\affiliation{\UT}
\author{Hiroyuki Takeya}
\affiliation{\NIMS}
\author{Yoshihiko Takano}
\affiliation{\NIMS}
\affiliation{\UT}

\date{\today}

\maketitle


\newpage
\section{Full Results of XERUS}
\begin{table}[H]
   \begin{ruledtabular}
   \begin{tabular}{ccccccc}
   
    Rank &                               Formula &                Spacegroup &               Provider &                            ID & Rwp (\%) &            Wt. Pct. (\%) \\\hline
   
       0 & \makecell{Ni \\ Cr$_{1.6}$Ni$_{0.4}$} & \makecell{\textit{Fm}$\bar{3}$\textit{m} \\ \textit{Im}$\bar{3}$\textit{m}} &  \makecell{COD \\ COD} & \makecell{2100646 \\ 1525376} &    26.01 & \makecell{65.75 \\34.25} \\\hline
       1 &                   \makecell{Ni \\ Cr} & \makecell{\textit{Fm}$\bar{3}$\textit{m} \\ \textit{Im}$\bar{3}$\textit{m}} &   \makecell{COD \\ MP} &   \makecell{2100646 \\ mp-90} &    27.35 & \makecell{67.93 \\32.07} \\\hline
       2 &                   \makecell{Ni \\ Cr} & \makecell{\textit{Fm}$\bar{3}$\textit{m} \\ \textit{Fm}$\bar{3}$\textit{m}} & \makecell{COD \\ OQMD} &  \makecell{2100646 \\ 592135} &    33.63 &  \makecell{95.34 \\4.66} \\\hline
       3 &                                    Ni &                     \textit{Fm}$\bar{3}$\textit{m} &                    COD &                       2100646 &    36.16 &                   100.00 \\\hline
       4 &                                    Cr &                     \textit{Im}$\bar{3}$\textit{m} &                     MP &                         mp-90 &    38.15 &                   100.00 \\\hline
       5 &                  Cr$_{1.6}$Ni$_{0.4}$ &                     \textit{Im}$\bar{3}$\textit{m} &                    COD &                       1525376 &    38.84 &                   100.00 \\\hline
   
   \end{tabular}
         
   \end{ruledtabular}
   \caption{XERUS results for Ni and Cr powder mixture for \textit{n\_runs} = 2, \textit{g} = 3 and $\delta$ = 1.3}
   \end{table}
   
\begin{table}[H]
   \begin{ruledtabular}
   \begin{tabular}{ccccccc}
   
    Rank &                            Formula &                  Spacegroup &              Provider &                               ID & Rwp (\%) &             Wt. Pct. (\%) \\\hline
   
       0 &  \makecell{HoB$_{2}$ \\ HoB$_{4}$} & \makecell{\textit{P}6/\textit{mmm} \\ \textit{P}4/\textit{mbm}} &   \makecell{MP \\ MP} &  \makecell{mp-2267 \\ mp-569281} &    23.53 &  \makecell{89.55 \\10.45} \\\hline
       1 &         \makecell{HoB$_{2}$ \\ Ho} &  \makecell{\textit{P}6/\textit{mmm} \\ \textit{Fm}$\bar{3}$\textit{m}} & \makecell{MP \\ OQMD} &      \makecell{mp-2267 \\ 10105} &    26.19 &   \makecell{96.72 \\3.28} \\\hline
       2 & \makecell{HoB$_{2}$ \\ HoB$_{12}$} &  \makecell{\textit{P}6/\textit{mmm} \\ \textit{Fm}$\bar{3}$\textit{m}} &   \makecell{MP \\ MP} & \makecell{mp-2267 \\ mp-1104585} &    26.66 &   \makecell{97.79 \\2.21} \\\hline
       3 &         \makecell{HoB$_{2}$ \\ Ho} &   \makecell{\textit{P}6/\textit{mmm} \\ \textit{R}$\bar{3}$\textit{m}} &   \makecell{MP \\ MP} &   \makecell{mp-2267 \\ mp-10659} &    26.89 & \makecell{100.32 \\-0.32} \\\hline
       4 &                          HoB$_{2}$ &                      \textit{P}6/\textit{mmm} &                    MP &                          mp-2267 &    29.46 &                    100.00 \\\hline
       5 &                                 Ho &                       \textit{Fm}$\bar{3}$\textit{m} &                  OQMD &                            10105 &    67.08 &                    100.00 \\\hline
       6 &                                 Ho &                        \textit{R}$\bar{3}$\textit{m} &                    MP &                         mp-10659 &    68.17 &                    100.00 \\\hline
   
   \end{tabular}
         
   \end{ruledtabular}
   \caption{XERUS results for a HoB$_{2.1}$ alloy for \textit{n\_runs} = 2, \textit{g} = 3 and $\delta$ = 1.3}
   \end{table}
\begin{table}[H]
   \begin{ruledtabular}
   \begin{tabular}{ccccccc}
   
    Rank &                                                                                    Formula &                    Spacegroup &               Provider &                            ID & Rwp (\%) &            Wt. Pct. (\%) \\\hline
   
       0 &                        \makecell{Sr$_{0.3}$La$_{3.7}$Cu$_{2}$O$_{7.2}$ \\ La$_{2}$O$_{3}$} &    \makecell{\textit{I}4/\textit{mmm} \\ \textit{P}$\bar{3}$\textit{m}1} &   \makecell{COD \\ MP} & \makecell{1529795 \\ mp-1968} &    18.16 &  \makecell{93.30 \\6.70} \\\hline
       1 &                        \makecell{Sr$_{0.3}$La$_{3.7}$Cu$_{2}$O$_{7.2}$ \\ La$_{2}$O$_{3}$} & \makecell{\textit{I}4/\textit{mmm} \\ \textit{P}6$_{3}$/\textit{mmc}} &  \makecell{COD \\ COD} & \makecell{1529795 \\ 2002286} &    18.17 &  \makecell{93.05 \\6.94} \\\hline
       2 &                        \makecell{Sr$_{0.3}$La$_{3.7}$Cu$_{2}$O$_{7.2}$ \\ La$_{2}$O$_{3}$} &     \makecell{\textit{I}4/\textit{mmm} \\ \textit{P}321} &  \makecell{COD \\ COD} & \makecell{1529795 \\ 1010278} &    18.42 &  \makecell{94.06 \\5.95} \\\hline
       3 &                      \makecell{Sr$_{0.3}$La$_{3.7}$Cu$_{2}$O$_{7.2}$ \\ La$_{2}$CuO$_{4}$} &   \makecell{\textit{I}4/\textit{mmm} \\ \textit{I}4/\textit{mmm}} & \makecell{COD \\ OQMD} &    \makecell{1529795 \\ 3754} &    19.42 &  \makecell{90.79 \\9.21} \\\hline
       4 & \makecell{Sr$_{0.3}$La$_{3.7}$Cu$_{2}$O$_{7.2}$ \\ Sr$_{0.52}$La$_{7.48}$Cu$_{4}$O$_{16}$} &     \makecell{\textit{I}4/\textit{mmm} \\ \textit{Cmce}} &  \makecell{COD \\ COD} & \makecell{1529795 \\ 2002462} &    20.55 & \makecell{41.70 \\58.30} \\\hline
       5 &                                                      Sr$_{0.3}$La$_{3.7}$Cu$_{2}$O$_{7.2}$ &                        \textit{I}4/\textit{mmm} &                    COD &                       1529795 &    21.86 &                   100.00 \\\hline
       6 &                                                     Sr$_{0.52}$La$_{7.48}$Cu$_{4}$O$_{16}$ &                          \textit{Cmce} &                    COD &                       2002462 &    24.01 &                   100.00 \\\hline
       7 &                                                                          La$_{2}$CuO$_{4}$ &                        \textit{I}4/\textit{mmm} &                   OQMD &                          3754 &    34.07 &                   100.00 \\\hline
   
   \end{tabular}
   \end{ruledtabular}
   \caption{XERUS results for a La$_{2-x}$Sr$_{x}$CuO$_{4}$ sample with $x$ = 0.25 for \textit{n\_runs} = 2, \textit{g} = 3 and $\delta$ = 1.3}
   \end{table}
   
      \setlength\tabcolsep{10pt}
\begin{longtable}[l]{ccccccc}
   \hline
   \hline
   
   Rank &                                                               Formula &                              Spacegroup &                       Provider &                                       ID & Rwp (\%) &                     Wt. Pct. (\%) \\\hline

   0 &             \makecell{Mo$_{8}$C$_{4.08}$ \\ HoMoC$_{2}$ \\ HoC$_{2}$} &       \makecell{\textit{Pbcn} \\ \textit{Pnma} \\ \textit{I}4/\textit{mmm}} & \makecell{COD \\ OQMD \\ OQMD} &      \makecell{1527881 \\ 28872 \\ 3668} &    15.17 &   \makecell{45.02 \\45.34 \\9.64} \\\hline
   1 &       \makecell{Mo$_{8}$C$_{4.08}$ \\ HoMoC$_{2}$ \\ Ho$_{4}$C$_{7}$} &       \makecell{\textit{Pbcn} \\ \textit{Pnma} \\ \textit{P}2$_{1}$/\textit{c}} &   \makecell{COD \\ OQMD \\ MP} &  \makecell{1527881 \\ 28872 \\ mp-15177} &    16.15 &  \makecell{44.93 \\43.10 \\11.96} \\\hline
   2 &             \makecell{Mo$_{8}$C$_{4.08}$ \\ Mo$_{2}$C \\ HoMoC$_{2}$} &        \makecell{\textit{Pbcn} \\ \textit{P}$\bar{3}$1\textit{m} \\ \textit{Pnma}} & \makecell{COD \\ OQMD \\ OQMD} &    \makecell{1527881 \\ 677963 \\ 28872} &    16.72 &  \makecell{50.71 \\-2.27 \\51.56} \\\hline
   3 &                   \makecell{Mo$_{8}$C$_{4.08}$ \\ MoC \\ HoMoC$_{2}$} &     \makecell{\textit{Pbcn} \\ \textit{P}6$_{3}$/\textit{mmc} \\ \textit{Pnma}} & \makecell{COD \\ OQMD \\ OQMD} &     \makecell{1527881 \\ 28887 \\ 28872} &    16.84 &  \makecell{50.78 \\-1.63 \\50.84} \\\hline
   4 &                    \makecell{Mo$_{8}$C$_{4.08}$ \\ HoMoC$_{2}$ \\ Ho} &     \makecell{\textit{Pbcn} \\ \textit{Pnma} \\ \textit{P}6$_{3}$/\textit{mmc}} &   \makecell{COD \\ OQMD \\ MP} &    \makecell{1527881 \\ 28872 \\ mp-144} &    16.85 &   \makecell{49.81 \\48.64 \\1.55} \\\hline
   5 &                          \makecell{Mo$_{8}$C$_{4.08}$ \\ HoMoC$_{2}$} &                 \makecell{\textit{Pbcn} \\ \textit{Pnma}} &         \makecell{COD \\ OQMD} &              \makecell{1527881 \\ 28872} &    16.89 &          \makecell{50.19 \\49.81} \\\hline
   6 &                \makecell{Mo$_{8}$C$_{4.08}$ \\ Ho \\ Ho$_{4}$C$_{7}$} &   \makecell{\textit{Pbcn} \\ \textit{P}6$_{3}$/\textit{mmc} \\ \textit{P}2$_{1}$/\textit{c}} &     \makecell{COD \\ MP \\ MP} & \makecell{1527881 \\ mp-144 \\ mp-15177} &    23.23 &   \makecell{59.07 \\7.90 \\33.03} \\\hline
   7 &                      \makecell{Mo$_{8}$C$_{4.08}$ \\ Ho \\ HoC$_{2}$} &   \makecell{\textit{Pbcn} \\ \textit{P}6$_{3}$/\textit{mmc} \\ \textit{I}4/\textit{mmm}} &   \makecell{COD \\ MP \\ OQMD} &     \makecell{1527881 \\ mp-144 \\ 3668} &    24.29 &  \makecell{74.32 \\11.19 \\14.50} \\\hline
   8 &               \makecell{Mo$_{8}$C$_{4.08}$ \\ MoC \\ Ho$_{4}$C$_{7}$} &   \makecell{\textit{Pbcn} \\ \textit{P}6$_{3}$/\textit{mmc} \\ \textit{P}2$_{1}$/\textit{c}} &   \makecell{COD \\ OQMD \\ MP} &  \makecell{1527881 \\ 28887 \\ mp-15177} &    24.42 &   \makecell{64.90 \\6.37 \\28.73} \\\hline
   9 &                     \makecell{Mo$_{8}$C$_{4.08}$ \\ MoC \\ HoC$_{2}$} &   \makecell{\textit{Pbcn} \\ \textit{P}6$_{3}$/\textit{mmc} \\ \textit{I}4/\textit{mmm}} & \makecell{COD \\ OQMD \\ OQMD} &      \makecell{1527881 \\ 28887 \\ 3668} &    24.69 &   \makecell{76.19 \\8.86 \\14.96} \\\hline
  10 &                            \makecell{Mo$_{8}$C$_{4.08}$ \\ MoC \\ Ho} & \makecell{\textit{Pbcn} \\ \textit{P}6$_{3}$/\textit{mmc} \\ \textit{P}6$_{3}$/\textit{mmc}} &   \makecell{COD \\ OQMD \\ MP} &    \makecell{1527881 \\ 28887 \\ mp-144} &    24.85 &   \makecell{79.62 \\8.52 \\11.87} \\\hline
  11 &          \makecell{Mo$_{8}$C$_{4.08}$ \\ Mo$_{7.72}$C$_{3.84}$ \\ Ho} &   \makecell{\textit{Pbcn} \\ \textit{Pca}2$_{1}$ \\ \textit{P}6$_{3}$/\textit{mmc}} &    \makecell{COD \\ COD \\ MP} &  \makecell{1527881 \\ 1536525 \\ mp-144} &    24.85 &  \makecell{-5.09 \\90.30 \\14.79} \\\hline
  12 &                      \makecell{Mo$_{8}$C$_{4.08}$ \\ Mo$_{2}$C \\ Ho} &    \makecell{\textit{Pbcn} \\ \textit{P}$\bar{3}$1\textit{m} \\ \textit{P}6$_{3}$/\textit{mmc}} &   \makecell{COD \\ OQMD \\ MP} &   \makecell{1527881 \\ 677963 \\ mp-144} &    25.00 &  \makecell{89.78 \\-3.48 \\13.70} \\\hline
  13 &                                   \makecell{Mo$_{8}$C$_{4.08}$ \\ Ho} &             \makecell{\textit{Pbcn} \\ \textit{P}6$_{3}$/\textit{mmc}} &           \makecell{COD \\ MP} &             \makecell{1527881 \\ mp-144} &    25.10 &          \makecell{86.89 \\13.11} \\\hline
  14 &         \makecell{Mo$_{8}$C$_{4.08}$ \\ Mo$_{7.72}$C$_{3.84}$ \\ MoC} &   \makecell{\textit{Pbcn} \\ \textit{Pca}2$_{1}$ \\ \textit{P}6$_{3}$/\textit{mmc}} &  \makecell{COD \\ COD \\ OQMD} &   \makecell{1527881 \\ 1536525 \\ 28887} &    25.37 &   \makecell{-5.57 \\95.78 \\9.78} \\\hline
  15 &                \makecell{Mo$_{8}$C$_{4.08}$ \\ Mo$_{7.72}$C$_{3.84}$} &               \makecell{\textit{Pbcn} \\ \textit{Pca}2$_{1}$} &          \makecell{COD \\ COD} &            \makecell{1527881 \\ 1536525} &    25.58 &         \makecell{-6.50 \\106.50} \\\hline
  16 &                                  \makecell{Mo$_{8}$C$_{4.08}$ \\ MoC} &             \makecell{\textit{Pbcn} \\ \textit{P}6$_{3}$/\textit{mmc}} &         \makecell{COD \\ OQMD} &              \makecell{1527881 \\ 28887} &    25.58 &           \makecell{90.13 \\9.87} \\\hline
  17 &                     \makecell{Mo$_{8}$C$_{4.08}$ \\ Mo$_{2}$C \\ MoC} &    \makecell{\textit{Pbcn} \\ \textit{P}$\bar{3}$1\textit{m} \\ \textit{P}6$_{3}$/\textit{mmc}} & \makecell{COD \\ OQMD \\ OQMD} &    \makecell{1527881 \\ 677963 \\ 28887} &    25.60 &   \makecell{85.76 \\4.20 \\10.04} \\\hline
  18 &   \makecell{Mo$_{8}$C$_{4.08}$ \\ Mo$_{7.72}$C$_{3.84}$ \\ Mo$_{2}$C} &      \makecell{\textit{Pbcn} \\ \textit{Pca}2$_{1}$ \\ \textit{P}$\bar{3}$1\textit{m}} &  \makecell{COD \\ COD \\ OQMD} &  \makecell{1527881 \\ 1536525 \\ 677963} &    25.78 &    \makecell{94.34 \\4.65 \\1.01} \\\hline
  19 &                            \makecell{Mo$_{8}$C$_{4.08}$ \\ Mo$_{2}$C} &                \makecell{\textit{Pbcn} \\ \textit{P}$\bar{3}$1\textit{m}} &         \makecell{COD \\ OQMD} &             \makecell{1527881 \\ 677963} &    26.12 &          \makecell{87.14 \\12.86} \\\hline
  20 &                                                    Mo$_{8}$C$_{4.08}$ &                                    \textit{Pbcn} &                            COD &                                  1527881 &    26.38 &                            100.00 \\\hline
  21 &                                                 Mo$_{7.72}$C$_{3.84}$ &                                  \textit{Pca}2$_{1}$ &                            COD &                                  1536525 &    26.54 &                            100.00 \\\hline
  22 &                                                             Mo$_{2}$C &                                   \textit{P}$\bar{3}$1\textit{m} &                           OQMD &                                   677963 &    28.82 &                            100.00 \\\hline
  23 & \makecell{Mo$_{8}$C$_{4.08}$ \\ Mo$_{7.72}$C$_{3.84}$ \\ HoMoC$_{2}$} &       \makecell{\textit{Pbcn} \\ \textit{Pca}2$_{1}$ \\ \textit{Pnma}} &  \makecell{COD \\ COD \\ OQMD} &   \makecell{1527881 \\ 1536525 \\ 28872} &    30.96 & \makecell{-5.41 \\-5.50 \\110.91} \\\hline
 \hline
   \hline
\end{longtable}
TABLE SM4. XERUS results for an alloy where the targeted composition was HoMoC$_{2}$ for \textit{n\_runs} = 3, \textit{g} = 3 and $\delta$ = 1.3

      \setlength\tabcolsep{10pt}
\begin{longtable}[l]{ccccccc}
   \hline
   \hline
    Rank &                                                                            Formula &                            Spacegroup &                       Provider &                                        ID & Rwp (\%) &                    Wt. Pct. (\%) \\\hline
   
       0 &                       \makecell{LiF \\ Li$_{36}$Mn$_{72}$O$_{144}$ \\ LiMnO$_{2}$} &  \makecell{\textit{Fm}$\bar{3}$\textit{m} \\ \textit{I}4$_{1}$/\textit{amd} \\ \textit{Pmmn}} &   \makecell{COD \\ COD \\ COD} &  \makecell{1010990 \\ 4001894 \\ 7214212} &     8.31 & \makecell{50.81 \\25.64 \\23.55} \\\hline
       1 &                 \makecell{LiF \\ Li$_{8.832}$Mn$_{15.168}$O$_{32}$ \\ LiMnO$_{2}$} &     \makecell{\textit{Fm}$\bar{3}$\textit{m} \\ \textit{Fd}$\bar{3}$\textit{m} \\ \textit{Pmmn}} &   \makecell{COD \\ COD \\ COD} &  \makecell{1010990 \\ 1514031 \\ 7214212} &     8.33 & \makecell{50.96 \\25.67 \\23.37} \\\hline
       2 &                                         \makecell{LiF \\ MnO$_{2}$ \\ LiMnO$_{2}$} &     \makecell{\textit{Fm}$\bar{3}$\textit{m} \\ \textit{Fd}$\bar{3}$\textit{m} \\ \textit{Pmmn}} &    \makecell{COD \\ MP \\ COD} & \makecell{1010990 \\ mp-25275 \\ 7214212} &    10.45 & \makecell{50.81 \\26.10 \\23.09} \\\hline
       3 &                          \makecell{LiF \\ Mn \\ Li$_{8.832}$Mn$_{15.168}$O$_{32}$} &    \makecell{\textit{Fm}$\bar{3}$\textit{m} \\ \textit{I}$\bar{4}$3\textit{m} \\ \textit{Fd}$\bar{3}$\textit{m}} &  \makecell{COD \\ OQMD \\ COD} &     \makecell{1010990 \\ 7874 \\ 1514031} &    11.70 &  \makecell{60.24 \\8.53 \\31.23} \\\hline
       4 &                                \makecell{LiF \\ Mn \\ Li$_{36}$Mn$_{72}$O$_{144}$} & \makecell{\textit{Fm}$\bar{3}$\textit{m} \\ \textit{I}$\bar{4}$3\textit{m} \\ \textit{I}4$_{1}$/\textit{amd}} &  \makecell{COD \\ OQMD \\ COD} &     \makecell{1010990 \\ 7874 \\ 4001894} &    11.73 &  \makecell{60.52 \\8.43 \\31.05} \\\hline
       5 &                                \makecell{LiF \\ Li$_{36}$Mn$_{72}$O$_{144}$ \\ Li} & \makecell{\textit{Fm}$\bar{3}$\textit{m} \\ \textit{I}4$_{1}$/\textit{amd} \\ \textit{I}$\bar{4}$3\textit{d}} &  \makecell{COD \\ COD \\ OQMD} &    \makecell{1010990 \\ 4001894 \\ 18968} &    12.10 & \makecell{46.95 \\20.36 \\32.69} \\\hline
       6 &                          \makecell{LiF \\ Mn \\ Li$_{8.832}$Mn$_{15.168}$O$_{32}$} &    \makecell{\textit{Fm}$\bar{3}$\textit{m} \\ \textit{Fm}$\bar{3}$\textit{m} \\ \textit{Fd}$\bar{3}$\textit{m}} &  \makecell{COD \\ OQMD \\ COD} &   \makecell{1010990 \\ 676149 \\ 1514031} &    12.12 &  \makecell{64.55 \\4.58 \\30.87} \\\hline
       7 &                          \makecell{LiF \\ Li$_{8.832}$Mn$_{15.168}$O$_{32}$ \\ Li} &    \makecell{\textit{Fm}$\bar{3}$\textit{m} \\ \textit{Fd}$\bar{3}$\textit{m} \\ \textit{I}$\bar{4}$3\textit{d}} &  \makecell{COD \\ COD \\ OQMD} &    \makecell{1010990 \\ 1514031 \\ 18968} &    12.14 & \makecell{46.96 \\20.71 \\32.33} \\\hline
       8 &                                \makecell{LiF \\ Mn \\ Li$_{36}$Mn$_{72}$O$_{144}$} & \makecell{\textit{Fm}$\bar{3}$\textit{m} \\ \textit{Fm}$\bar{3}$\textit{m} \\ \textit{I}4$_{1}$/\textit{amd}} &  \makecell{COD \\ OQMD \\ COD} &   \makecell{1010990 \\ 676149 \\ 4001894} &    12.24 &  \makecell{64.74 \\4.46 \\30.80} \\\hline
       9 &                 \makecell{LiF \\ Li$_{36}$Mn$_{72}$O$_{144}$ \\ LiMn$_{2}$O$_{4}$} &  \makecell{\textit{Fm}$\bar{3}$\textit{m} \\ \textit{I}4$_{1}$/\textit{amd} \\ \textit{Fddd}} &   \makecell{COD \\ COD \\ COD} &  \makecell{1010990 \\ 4001894 \\ 1514020} &    12.56 & \makecell{72.48 \\33.59 \\-6.07} \\\hline
      10 &                         \makecell{LiF \\ Li$_{36}$Mn$_{72}$O$_{144}$ \\ MnO$_{2}$} & \makecell{\textit{Fm}$\bar{3}$\textit{m} \\ \textit{I}4$_{1}$/\textit{amd} \\ \textit{Fd}$\bar{3}$\textit{m}} &    \makecell{COD \\ COD \\ MP} & \makecell{1010990 \\ 4001894 \\ mp-25275} &    12.62 &  \makecell{68.91 \\26.43 \\4.66} \\\hline
      11 &           \makecell{LiF \\ Li$_{8.832}$Mn$_{15.168}$O$_{32}$ \\ LiMn$_{2}$O$_{4}$} &     \makecell{\textit{Fm}$\bar{3}$\textit{m} \\ \textit{Fd}$\bar{3}$\textit{m} \\ \textit{Fddd}} &   \makecell{COD \\ COD \\ COD} &  \makecell{1010990 \\ 1514031 \\ 1514020} &    12.67 &  \makecell{68.74 \\28.48 \\2.78} \\\hline
      12 & \makecell{LiF \\ Li$_{8.832}$Mn$_{15.168}$O$_{32}$ \\ Li$_{36}$Mn$_{72}$O$_{144}$} & \makecell{\textit{Fm}$\bar{3}$\textit{m} \\ \textit{Fd}$\bar{3}$\textit{m} \\ \textit{I}4$_{1}$/\textit{amd}} &   \makecell{COD \\ COD \\ COD} &  \makecell{1010990 \\ 1514031 \\ 4001894} &    12.70 &  \makecell{69.39 \\21.33 \\9.28} \\\hline
      13 &                   \makecell{LiF \\ Li$_{8.832}$Mn$_{15.168}$O$_{32}$ \\ MnO$_{2}$} &    \makecell{\textit{Fm}$\bar{3}$\textit{m} \\ \textit{Fd}$\bar{3}$\textit{m} \\ \textit{Fd}$\bar{3}$\textit{m}} &    \makecell{COD \\ COD \\ MP} & \makecell{1010990 \\ 1514031 \\ mp-25275} &    12.71 &  \makecell{68.74 \\26.77 \\4.49} \\\hline
      14 &                                      \makecell{LiF \\ Li$_{36}$Mn$_{72}$O$_{144}$} &          \makecell{\textit{Fm}$\bar{3}$\textit{m} \\ \textit{I}4$_{1}$/\textit{amd}} &          \makecell{COD \\ COD} &             \makecell{1010990 \\ 4001894} &    12.73 &         \makecell{69.77 \\30.23} \\\hline
      15 &                                \makecell{LiF \\ Li$_{8.832}$Mn$_{15.168}$O$_{32}$} &             \makecell{\textit{Fm}$\bar{3}$\textit{m} \\ \textit{Fd}$\bar{3}$\textit{m}} &          \makecell{COD \\ COD} &             \makecell{1010990 \\ 1514031} &    12.74 &         \makecell{69.45 \\30.55} \\\hline
      16 &                                   \makecell{LiF \\ MnO$_{2}$ \\ LiMn$_{2}$O$_{4}$} &     \makecell{\textit{Fm}$\bar{3}$\textit{m} \\ \textit{Fd}$\bar{3}$\textit{m} \\ \textit{Fddd}} &    \makecell{COD \\ MP \\ COD} & \makecell{1010990 \\ mp-25275 \\ 1514020} &    13.06 &  \makecell{59.93 \\5.68 \\34.38} \\\hline
      17 &                                                  \makecell{LiF \\ Mn \\ MnO$_{2}$} &    \makecell{\textit{Fm}$\bar{3}$\textit{m} \\ \textit{I}$\bar{4}$3\textit{m} \\ \textit{Fd}$\bar{3}$\textit{m}} &   \makecell{COD \\ OQMD \\ MP} &    \makecell{1010990 \\ 7874 \\ mp-25275} &    13.20 &  \makecell{59.48 \\8.62 \\31.91} \\\hline
      18 &                                                  \makecell{LiF \\ MnO$_{2}$ \\ Li} &    \makecell{\textit{Fm}$\bar{3}$\textit{m} \\ \textit{Fd}$\bar{3}$\textit{m} \\ \textit{I}$\bar{4}$3\textit{d}} &   \makecell{COD \\ MP \\ OQMD} &   \makecell{1010990 \\ mp-25275 \\ 18968} &    13.54 & \makecell{47.10 \\21.31 \\31.59} \\\hline
      19 &                                                  \makecell{LiF \\ Mn \\ MnO$_{2}$} &    \makecell{\textit{Fm}$\bar{3}$\textit{m} \\ \textit{Fm}$\bar{3}$\textit{m} \\ \textit{Fd}$\bar{3}$\textit{m}} &   \makecell{COD \\ OQMD \\ MP} &  \makecell{1010990 \\ 676149 \\ mp-25275} &    13.54 &  \makecell{63.92 \\4.57 \\31.51} \\\hline
      20 &                                                        \makecell{LiF \\ MnO$_{2}$} &             \makecell{\textit{Fm}$\bar{3}$\textit{m} \\ \textit{Fd}$\bar{3}$\textit{m}} &           \makecell{COD \\ MP} &            \makecell{1010990 \\ mp-25275} &    14.05 &         \makecell{68.85 \\31.15} \\\hline
      21 &                                                         \makecell{LiF \\ Mn \\ Mn} &    \makecell{\textit{Fm}$\bar{3}$\textit{m} \\ \textit{I}$\bar{4}$3\textit{m} \\ \textit{Fm}$\bar{3}$\textit{m}} & \makecell{COD \\ OQMD \\ OQMD} &      \makecell{1010990 \\ 7874 \\ 676149} &    18.73 &   \makecell{87.17 \\9.27 \\3.56} \\\hline
      22 &                                                               \makecell{LiF \\ Mn} &             \makecell{\textit{Fm}$\bar{3}$\textit{m} \\ \textit{I}$\bar{4}$3\textit{m}} &         \makecell{COD \\ OQMD} &                \makecell{1010990 \\ 7874} &    18.78 &         \makecell{88.81 \\11.19} \\\hline
      23 &                                                               \makecell{LiF \\ Mn} &             \makecell{\textit{Fm}$\bar{3}$\textit{m} \\ \textit{Fm}$\bar{3}$\textit{m}} &         \makecell{COD \\ OQMD} &              \makecell{1010990 \\ 676149} &    19.02 &          \makecell{93.27 \\6.73} \\\hline
      24 &                                                                                LiF &                                 \textit{Fm}$\bar{3}$\textit{m} &                            COD &                                   1010990 &    19.32 &                           100.00 \\\hline
      25 &                                                                                 Mn &                                 \textit{I}$\bar{4}$3\textit{m} &                           OQMD &                                      7874 &    20.39 &                           100.00 \\\hline
      26 &                                                                                 Mn &                                 \textit{Fm}$\bar{3}$\textit{m} &                           OQMD &                                    676149 &    25.86 &                           100.00 \\\hline

   \end{longtable}
   Table SM5. XERUS results for the LiMn$_{2}$O$_{4}$+LiF+Li$_{2}$MnO$_{3}$ mixture for \textit{n\_runs} = \textit{auto}, \textit{g} = 3 and $\delta$ = 1.3

   \begin{table}[H]
      \begin{ruledtabular}
      \begin{tabular}{ccccccc}
      
       Rank &                                                                        Formula &                   Spacegroup &              Provider &                            ID & Rwp (\%) &            Wt. Pct. (\%) \\\hline
      
          0 &              \makecell{Li$_{8.832}$Mn$_{15.168}$O$_{32}$ \\ Li$_{2}$MnO$_{3}$} &     \makecell{\textit{Fd}$\bar{3}$\textit{m} \\ \textit{C}2/\textit{c}} & \makecell{COD \\ COD} & \makecell{1514031 \\ 1514065} &     7.08 & \makecell{49.81 \\50.19} \\\hline
          1 & \makecell{Li$_{8.832}$Mn$_{15.168}$O$_{32}$ \\ Li$_{7.98}$Mn$_{4.02}$O$_{12}$} &     \makecell{\textit{Fd}$\bar{3}$\textit{m} \\ \textit{C}2/\textit{m}} & \makecell{COD \\ COD} & \makecell{1514031 \\ 1544474} &     7.27 & \makecell{48.38 \\51.62} \\\hline
          2 &                             \makecell{Li$_{8.832}$Mn$_{15.168}$O$_{32}$ \\ Mn} &    \makecell{\textit{Fd}$\bar{3}$\textit{m} \\ \textit{I}$\bar{4}$3\textit{m}} &  \makecell{COD \\ MP} &   \makecell{1514031 \\ mp-35} &     8.66 & \makecell{78.14 \\21.86} \\\hline
          3 &              \makecell{Li$_{8.832}$Mn$_{15.168}$O$_{32}$ \\ LiMn$_{2}$O$_{4}$} &     \makecell{\textit{Fd}$\bar{3}$\textit{m} \\ \textit{Fddd}} & \makecell{COD \\ COD} & \makecell{1514031 \\ 1514020} &    10.13 & \makecell{63.42 \\36.58} \\\hline
          4 &    \makecell{Li$_{8.832}$Mn$_{15.168}$O$_{32}$ \\ Li$_{36}$Mn$_{72}$O$_{144}$} & \makecell{\textit{Fd}$\bar{3}$\textit{m} \\ \textit{I}4$_{1}$/\textit{amd}} & \makecell{COD \\ COD} & \makecell{1514031 \\ 4001894} &    10.55 & \makecell{76.53 \\23.47} \\\hline
          5 &                                              Li$_{8.832}$Mn$_{15.168}$O$_{32}$ &                        \textit{Fd}$\bar{3}$\textit{m} &                   COD &                       1514031 &    10.61 &                   100.00 \\\hline
          6 &                                                    Li$_{36}$Mn$_{72}$O$_{144}$ &                     \textit{I}4$_{1}$/\textit{amd} &                   COD &                       4001894 &    11.49 &                   100.00 \\\hline
          7 &                                                              LiMn$_{2}$O$_{4}$ &                         \textit{Fddd} &                   COD &                       1514020 &    12.93 &                   100.00 \\\hline
      
      \end{tabular}
            
      \end{ruledtabular}
      \caption{XERUS results for the LiMn$_{2}$O$_{4}$+Li$_{2}$MnO$_{3}$ mixture for \textit{n\_runs} = \textit{auto}, \textit{g} = 3 and $\delta$ = 1.3}
      \end{table}